\documentclass[a4paper,12pt,reqno]{article}
\usepackage{amsmath,amsfonts,amssymb,amsthm,dsfont}

\usepackage{hyperref}
\allowdisplaybreaks
\numberwithin{equation}{section}

\newcommand{\com}[2]{[\,#1\, ,\,#2\,]}	

\newcommand{\Om}{\Omega}
\newcommand{\om}{\omega}
\newcommand{\half}{\tfrac{1}{2}}
\newcommand{\ihalf}{\tfrac{\mathrm{i}}{2}}
\newcommand{\ifourth}{\tfrac{\mathrm{i}}{4}}
\newcommand{\quart}{\tfrac{1}{4}}
\newcommand{\Ii}{\mathrm{i}}

\newcommand{\4}{\tilde}

\newcommand{\6}{\partial}
\newcommand{\7}{\hat}

\newcommand{\ua}{{\underline{\alpha}}}
\newcommand{\ub}{{\underline{\beta}}}
\newcommand{\ug}{{\underline{\gamma}}}
\newcommand{\ud}{{\underline{\delta}}}
\newcommand{\ul}[1]{{\underline{#1}}}
\newcommand{\ol}[1]{{\overline{#1}}}

\newcommand{\acom}[2]{\{#1\, ,\,#2\}}	

\newcommand{\gam}{\Gamma}	
\newcommand{\GAM}{\hat{\Gamma}}	
\newcommand{\CC}{C}	
\newcommand{\IC}{C^{-1}}	
\newcommand{\unit}{\mathds{1}}

\newcommand{\Dim}{D} 
\newcommand{\Ns}{N} 
\newcommand{\Omg}{\Omega_\mathrm{gh}} 
\newcommand{\hOmg}{\hat\Omega}
\newcommand{\fb}{s_\mathrm{gh}} 
\newcommand{\Hg}{H_\mathrm{gh}} 
\newcommand{\hHg}{\7H_\mathrm{gh}} 
\newcommand{\fbb}[1]{s_{\mathrm{gh},#1}} 
\newcommand{\cdeg}{$c$-degree} 
\newcommand{\ddc}[1]{\frac{\partial}{\partial \tilde{c}^{#1}}}
\newcommand{\dds}[1]{\frac{\partial}{\partial #1}}
\newcommand{\xx}[1]{\xi^{\ul{#1}}} 
\newcommand{\pb}[1]{\bar\psi^{#1}}
\newcommand{\cb}[1]{\bar\chi^{#1}}

\newcommand{\tc}[1]{\tilde{c}^{\,#1}}

\newcommand{\LRA}{\Leftrightarrow}
\newcommand{\Then}{\Rightarrow}

\newcommand{\so}[1]{$\mathfrak{so}(#1)$}
\newcommand{\pap}{\cite{Brandt:2009xv}}
\newcommand{\papII}{\cite{Brandt:2010fa}}
\newcommand{\QED}{\hfill$\blacksquare$}

\newtheorem{prop}{Proposition}[section]
\newtheorem{lemma}[prop]{Lemma}

\begin{document}

\begin{flushright}
ITP--UH--09/10
\end{flushright}

\begin{center}
 {\large\bfseries Supersymmetry algebra cohomology III:\\[6pt] Primitive elements in four and five dimensions}
 \\[5mm]
 Friedemann Brandt \\[2mm]
 \textit{Institut f\"ur Theoretische Physik, Leibniz Universit\"at Hannover, Appelstra\ss e 2, D-30167 Hannover, Germany}
\end{center}

\begin{abstract}
The primitive elements of the supersymmetry algebra cohomology as defined in a previous paper are computed for standard supersymmetry algebras in four and five dimensions, for all signatures of the metric and any number of supersymmetries.
\end{abstract}

\tableofcontents

\section{Introduction}

This paper relates to supersymmetry algebra cohomology as defined in \pap, for supersymmetry algebras in $\Dim=4$ and $\Dim=5$ dimensions of translational generators $P_a$ ($a=1,\dots,\Dim$) and supersymmetry generators $Q^i_\ua$ ($i=1,\dots,\Ns$, $\ua=\ul1,\dots,\ul4$) of the form
 \begin{align}
  \com{P_a}{P_b}=0,\quad \com{P_a}{Q^i_\ua}=0,\quad \acom{Q^i_\ua}{Q^j_\ub}=M^{ij}\,(\gam^a \IC)_{\ua\ub}P_a
	\label{i3-1} 
 \end{align}
where $M^{ij}$ are the entries of an $\Ns\times\Ns$ matrix $M$ given by
  \begin{align}
  \Dim=4:\ M=
      \begin{pmatrix} -\Ii & 0 & \cdots & 0\\ 
                      0 & -\Ii & \cdots & 0\\
                      \vdots & \vdots & \ddots & \vdots\\
                      0 & 0 & \cdots & -\Ii                
      \end{pmatrix},\quad
  \Dim=5:\ M=
      \begin{pmatrix} \sigma_2 & 0 & \cdots & 0\\ 
                      0 & \sigma_2 & \cdots & 0\\
                      \vdots & \vdots & \ddots & \vdots\\
                      0 & 0 & \cdots & \sigma_2                
      \end{pmatrix}
	\label{i3-2} 
 \end{align}
with $\Ii$ denoting the imaginary unit and $\sigma_2$ denoting the second Pauli-matrix (hence, in the $\Dim=5$ we consider $\Ns=2,4,6,\dots$).

The object of this paper is the determination of the primitive elements of the supersymmetry algebra cohomology for these supersymmetry algebras \eqref{i3-1} for all signatures $(t,\Dim-t)$ ($t=0,\dots,\Dim$) of the Clifford algebra of the gamma matrices $\gam^a$.
According to the definition given in \pap, these primitive elements are the representatives of the cohomology $\Hg(\fb)$ of the coboundary operator
 \begin{align}
  \fb=-\half\, M^{ij}\,(\gam^a \IC)_{\ua\ub}\,\xi^\ua_i\xi^\ub_j\,\frac{\6}{\6 c^a}
	\label{i3-6} 
 \end{align}
in the space $\Omg$ of polynomials in translation ghosts $c^a$ and supersymmetry ghosts $\xi^\ua_i$ corresponding to the translational generators $P_a$ and the supersymmetry generators $Q^i_\ua$, respectively,
 \begin{align}
  &\Omg=\Big\{\sum_{p=0}^\Dim\sum_{n=0}^r c^{a_1}\dots c^{a_p}\xi^{\ua_1}_{i_1}\dots\xi^{\ua_n}_{i_n}
  a^{i_1\dots i_n}_{\ua_1\dots\ua_n a_1\dots a_p}\,|\,
  a^{i_1\dots i_n}_{\ua_1\dots\ua_n a_1\dots a_p}\in\mathbb{C},\ r=0,1,2,\ldots\Big\}.
	\label{i3-7} 
 \end{align}

For signatures $(1,3)$, $(2,2)$ and $(3,1)$ in $\Dim=4$ and signatures $(2,3)$ and $(3,2)$ in $\Dim=5$ the supersymmetry ghosts are Majorana spinors, for signatures $(0,4)$ and $(4,0)$ in $\Dim=4$ and signatures $(0,5)$, $(1,4)$, $(4,1)$ and $(5,0)$ in $\Dim=5$ they are symplectic Majorana spinors, cf. sections 2 and 4 of \pap. We note that for signature $(2,2)$ in $\Dim=4$ each Majorana spinor consists of two Majorana-Weyl spinors of opposite chirality and for signatures $(0,4)$ and $(4,0)$ each symplectic Majorana spinor consists of two symplectic Majorana-Weyl spinors of opposite chirality. $\Ns$ denotes in all cases the number of Majorana or symplectic Majorana supersymmetry ghosts (hence, for signature $(2,2)$ and $\Ns=1$ one has one Majorana supersymmetry ghost and thus two Majorana-Weyl supersymmetry ghosts etc.). 

Analogously to the strategy applied in \papII\ in two and three dimensions, we shall first compute $\Hg(\fb)$ in $\Dim=4$ explicitly in a particular spinor representation and then covariantize the results to make them independent of the spinor representation. Afterwards the results in $\Dim=5$ are derived by means of the results in $\Dim=4$. The particular spinor representations are defined by
 \begin{align}
 &\gam_1=k_1\,\sigma_1\otimes\sigma_0\, ,\ \gam_2=k_2\,\sigma_2\otimes\sigma_0\, ,\ 
  \gam_3=k_3\,\sigma_3\otimes\sigma_1\, ,\ \gam_4=k_4\,\sigma_3\otimes\sigma_2\,,\notag\\
 &\Dim=4:\quad
  \GAM=\sigma_3\otimes\sigma_3\,,\quad
  C=\sigma_2\otimes\sigma_1\, ,\notag\\
 &\Dim=5:\quad
  \gam_5=k_5\,\sigma_3\otimes\sigma_3\,,\quad
  C=\sigma_1\otimes\sigma_2
  \label{4D1}
 \end{align}
with
 \begin{align}
  k_a=\left\{\begin{array}{rl}
  \Ii & \mathrm{for}\ a\leq t\\
  1& \mathrm{for}\ a>t\end{array}\right. .
  \label{4D1a}
 \end{align}

As in \papII\ we shall use the notation $\sim$ for equivalence in $\Hg(\fb)$, i.e. for $\om_1,\om_2\in\Omg$ the notation $\om_1\sim \om_2$ means $\om_1-\om_2=\fb\om_3$ for some $\om_3\in\Omg$:
\begin{align}
  \om_1\sim \om_2\quad :\Leftrightarrow \quad \exists\,\om_3:\ \om_1-\om_2=\fb\om_3\quad (\om_1,\om_2,\om_3\in\Omg).
  \label{equiv}
\end{align}

Furthermore the paper uses terminology,
notation and conventions introduced in \pap.

\section{Primitive elements in four dimensions}\label{4D.0}

\subsection{\texorpdfstring{$\Hg(\fb)$ in particular spinor representations}{H(gh) in particular spinor representations}}\label{4D.1}

We shall first compute $\Hg(\fb)$ for $\Dim=4$ in the particular spinor representations \eqref{4D1}.
In order to do this for all signatures $(t,4-t)$ at once
we introduce the following translation ghost variables:
 \begin{align}
 &\tc1=-\half(k_1\,c^1+\Ii\,k_2\,c^2)\, ,\quad \tc2=-\half(k_1\,c^1-\Ii\,k_2\,c^2)\, ,\notag\\ 
 &\tc3=-\half(k_3\,c^3+\Ii\,k_4\,c^4)\, ,\quad \tc4=\half(k_3\,c^3-\Ii\,k_4\,c^4)\, .
  \label{4D5}
 \end{align}
Furthermore, in order to simplify the notation, we denote the components of $\xi_i=(\xx1_i,\xx2_i,\xx3_i,\xx4_i)$ by $\psi_i$, $\pb{i}$, $-\cb{i}$ and $\chi_i$: 
\begin{align}
  \xi_i=(\xx1_i,\xx2_i,\xx3_i,\xx4_i)=(\psi_i,\pb{i},-\cb{i},\chi_i).
	\label{4D3} 
\end{align}
$\psi_i$ and $\chi_i$ are the components of Weyl spinors $\xi_i^+$ with positive chirality, $\pb{i}$ and $\cb{i}$ the components of Weyl spinors $\xi_i^-$ with negative chirality ($\xi_i^\pm\GAM=\pm\xi_i^\pm$),
\begin{align}
  \xi_i^+=(\psi_i,0,0,\chi_i),\quad \xi_i^-=(0,\pb{i},-\cb{i},0).
	\label{4D3a} 
\end{align}
In terms of the ghost variables \eqref{4D5} and \eqref{4D3}, the $\fb$-transformations of the translation ghost variables are, for all signatures $(t,4-t)$:
\begin{align}
  \fb \tc1=\psi_i\pb{i},\ \fb \tc2=\chi_i\cb{i},\ 
  \fb \tc3=\psi_i\cb{i},\ \fb \tc4=\chi_i\pb{i}.
	\label{4D6} 
\end{align}

\subsubsection{\texorpdfstring{$\Hg(\fb)$ for $\Ns=1$}{H(gh) for N=1}}\label{4D.3}

In order to determine $\Hg(\fb)$ in the case $\Ns=1$, we first compute the cohomology of $\fb$ in the space $\Om^-$ of polynomials in the ghost variables $\tc1,\tc2,\tc3,\tc4,\pb1,\cb1$,
\begin{align}
  \Om^-=\left\{\om\in\Omg\left|\ \frac{\6\om}{\6\psi_1}=0\ \wedge\ \frac{\6\om}{\6\chi_1}=0\right.\right\}.
	\label{4D12} 
\end{align}

\begin{lemma}
\label{lem4D1}\quad \\
(i) A polynomial $\om\in\Om^-$ is $\fb$-closed if and only if it is a polynomial in $\pb1$, $\cb1$, $\tc2\pb1-\tc4\cb1$ and $\tc1\cb1-\tc3\pb1$:
\begin{align}
  \om\in\Om^-:\ \fb\om=0\ \LRA\ 
  \om=&\,p_1(\pb1,\cb1)\notag\\
  &+(\tc2\pb1-\tc4\cb1)p_2(\pb1,\cb1)\notag\\
  &+(\tc1\cb1-\tc3\pb1)p_3(\pb1,\cb1)\notag\\
  &+(\tc1\cb1-\tc3\pb1)(\tc2\pb1-\tc4\cb1)p_4(\pb1,\cb1)
  \label{4D46}
\end{align}
with polynomials $p_1(\pb1,\cb1)$,\dots, $p_4(\pb1,\cb1)$ in $\pb1$ and $\cb1$.\\
(ii) No nonvanishing polynomial in $\pb1$, $\cb1$, $\tc2\pb1-\tc4\cb1$ and $\tc1\cb1-\tc3\pb1$ is a coboundary in $\Hg(\fb)$,
\begin{align}
  &p_1(\pb1,\cb1)+(\tc2\pb1-\tc4\cb1)p_2(\pb1,\cb1)
  +(\tc1\cb1-\tc3\pb1)p_3(\pb1,\cb1)\notag\\
  &+(\tc1\cb1-\tc3\pb1)(\tc2\pb1-\tc4\cb1)p_4(\pb1,\cb1)\sim 0
  \notag\\
  &\LRA\ \forall i\in\{1,2,3,4\}:\ p_i(\pb1,\cb1)=0.
  \label{4D46a}
\end{align}
\end{lemma}

{\bf Proof:}
We decompose an $\fb$-cocycle $\om\in\Om^-$ into parts $\om^p$ with definite \cdeg\ $p$. 
Since $\fb$ decrements the \cdeg\ by one unit, all parts $\om^p$ are $\fb$-cocycles,
\begin{align}
  \fb\om=0,\quad\om=\sum_{p=0}^4\om^p,\quad N_c\om^p=p\,\om^p\quad \Then\quad  \fb\om^p=0\quad\forall p 
	\label{4D13a} 
\end{align}
where $N_c$ denotes the counting operator for the translation ghosts,
\begin{align}
  N_c=c^a\,\dds{c^a}\, .
	\label{4D15} 
\end{align}
Hence, we can determine the $\fb$-cocycles in $\Om^-$ separately for the various \cdeg s. To determine these $\fb$-cocycles, we use that
$\fb$ acts in terms of the ghost variables $\tc1,\tc2,\tc3,\tc4,\psi_1,\pb1,\chi_1,\cb1$ according to
\begin{align}
  \fb=\psi_1\pb{1}\ddc1+\chi_1\cb{1}\ddc2+\psi_1\cb{1}\ddc3+\chi_1\pb{1}\ddc4
  =\psi_1 D_1+\chi_1 D_2
	\label{4D10} 
\end{align}
with
\begin{align}
  D_1=\pb{1}\ddc1+\cb{1}\ddc3 \, ,\quad D_2=\cb{1}\ddc2+\pb{1}\ddc4\, .
	\label{4D11} 
\end{align}
As an element $\om$ of $\Om^-$ neither depends on $\psi_1$ nor on $\chi_1$ and as $D_1$ and $D_2$ do not involve $\psi_1$ or $\chi_1$, the cocycle condition $\fb\om=(\psi_1D_1+\chi_1D_2)\om=0$ imposes $D_1\om=0$ and $D_2\om=0$. Accordingly, any $\fb$-cocycle in $\Om^-$ with \cdeg\ $p$ is annihiliated both by $D_1$ and $D_2$:
\begin{align}
  \fb\om^p=0,\ \om^p\in\Om^-\quad\LRA\quad D_1\om^p=0\ \wedge\ D_2\om^p=0. 
	\label{4D12a} 
\end{align}
Part (i) of the lemma is obtained easily by solving $D_1\om^p=D_2\om^p=0$ directly for the various values of $p$. We shall explicitly spell out the computation only for the most involved case $p=2$.
In this case we have
\begin{align}
  \om^2=\tc1\tc2 f_{12}+\tc1\tc3 f_{13}+\tc1\tc4 f_{14}+\tc2\tc3 f_{23}+\tc2\tc4 f_{24}+\tc3\tc4 f_{34} 
	\label{4D12b} 
\end{align}
with $f_{ij}=f_{ij}(\cb{1},\pb{1})$. Applying $D_1$ to $\om^2$ yields
\begin{align}
  D_1\om^2=\tc2 (\pb{1}f_{12}-\cb{1}f_{23})+(\tc3\pb{1}-\tc1\cb{1})f_{13}+\tc4 (\pb{1}f_{14}+\cb{1}f_{34}).
	\label{4D12c} 
\end{align}
$D_1\om^2=0$ imposes thus
\[
\pb{1}f_{12}=\cb{1}f_{23},\quad f_{13}=0,\quad \pb{1}f_{14}=-\cb{1}f_{34}
\]
which implies
\begin{align}
  f_{12}=\cb{1}g_1, \quad f_{23}=\pb{1}g_1,\quad f_{13}=0,\quad
  f_{14}=\cb{1}g_2,\quad f_{34}=-\pb{1}g_2
	\label{4D12d} 
\end{align}
for some $g_i=g_i(\cb{1},\pb{1})$.
Using \eqref{4D12d} in \eqref{4D12b}, we obtain the intermediate result
\begin{align}
  \om^2=(\tc1\cb{1}-\tc3 \pb{1})(\tc2 g_1+\tc4 g_2)+\tc2\tc4 f_{24}. 
	\label{4D12dd} 
\end{align}
Applying now $D_2$ to \eqref{4D12dd} yields
\begin{align}
  D_2\om^2=-(\tc1\cb{1}-\tc3 \pb{1})(\cb{1} g_1+\pb{1} g_2)+(\tc4\cb{1}-\tc2\pb{1}) f_{24}.
	\label{4D12e} 
\end{align}
$D_2\om^2=0$ thus imposes 
\[
\cb{1} g_1=-\pb{1} g_2,\quad f_{24}=0
\]
which implies
\begin{align}
  g_1=\pb{1}p_4,\quad g_2=-\cb{1}p_4,\quad f_{24}=0  
	\label{4D12f} 
\end{align}
for some $p_4=p_4(\pb1,\cb1)$. Using \eqref{4D12f} in \eqref{4D12dd}, we conclude
\begin{align}
  \om^2=(\tc1\cb{1}-\tc3 \pb{1})(\tc2 \pb{1}-\tc4\cb{1})p_4(\pb1,\cb1) 
	\label{4D12ff} 
\end{align}
which provides the last contribution to $\om$ in \eqref{4D46}.

Analogously one derives for $p=0,3,4,1$, respectively:
\begin{align}
&\om^0=p_1(\pb1,\cb1),\quad \om^3=0,\quad \om^4=0,\notag \\
&\om^1=(\tc2\pb1-\tc4\cb1)p_2(\pb1,\cb1)
  +(\tc1\cb1-\tc3\pb1)p_3(\pb1,\cb1).\label{4D12g} 
\end{align}
\eqref{4D12ff} and \eqref{4D12g} provide part (i) of lemma \ref{lem4D1}. Part (ii) of the lemma holds because cocycles of $\fb$ which neither depend on $\psi_1$ nor on $\chi_1$ cannot be exact in $\Hg(\fb)$ since every term in \eqref{4D10} is linear in $\psi_1$ or $\chi_1$ (coboundaries contain only terms that depend at least linearly on $\psi_1$ or $\chi_1$). This completes the proof of lemma \ref{lem4D1}. \QED

A result analogous to lemma \ref{lem4D1} holds for the cohomology of $\fb$ in the space $\Om^+$ of polynomials in the ghost variables $\tc1,\tc2,\tc3,\tc4,\psi_1,\chi_1$,
\begin{align}
  \Om^+=\left\{\om\in\Omg\left|\ \frac{\6\om}{\6\pb1}=0\ \wedge\ \frac{\6\om}{\6\cb1}=0\right.\right\}.
	\label{4D47} 
\end{align}

\begin{lemma}
\label{lem4D2}\quad \\
(i) A polynomial $\om\in\Om^+$ is $\fb$-closed if and only if it is a polynomial in $\psi_1$, $\chi_1$, $\tc2\psi_1-\tc3\chi_1$ and $\tc1\chi_1-\tc4\psi_1$:
\begin{align}
  \om\in\Om^+:\ \fb\om=0\ \LRA\ 
  \om=&\,q_1(\psi_1,\chi_1)\notag\\
  &+(\tc2\psi_1-\tc3\chi_1)q_2(\psi_1,\chi_1)\notag\\
  &+(\tc1\chi_1-\tc4\psi_1)q_3(\psi_1,\chi_1)\notag\\
  &+(\tc1\chi_1-\tc4\psi_1)(\tc2\psi_1-\tc3\chi_1)q_4(\psi_1,\chi_1)
  \label{4D48}
\end{align}
with polynomials $q_1(\psi_1,\chi_1)$,\dots,$q_4(\psi_1,\chi_1)$ in $\psi_1$ and $\chi_1$.\\
(ii) No nonvanishing polynomial in $\psi_1$, $\chi_1$, $\tc2\psi_1-\tc3\chi_1$ and $\tc1\chi_1-\tc4\psi_1$ is a coboundary in $\Hg(\fb)$,
\begin{align}
  &q_1(\psi_1,\chi_1)+(\tc2\psi_1-\tc3\chi_1)q_2(\psi_1,\chi_1)
  +(\tc1\chi_1-\tc4\psi_1)q_3(\psi_1,\chi_1)\notag\\
  &+(\tc1\chi_1-\tc4\psi_1)(\tc2\psi_1-\tc3\chi_1)q_4(\psi_1,\chi_1)\sim 0\notag\\
  &\LRA\ \forall i\in\{1,2,3,4\}:\ q_i(\psi_1,\chi_1)=0.
  \label{4D48a}
\end{align}
\end{lemma}

Lemmas \ref{lem4D1} and \ref{lem4D2} yield all those cocycles of $\Hg(\fb)$ which do not depend either on $\psi_1$ and $\chi_1$ or on $\pb1$ and $\cb1$. The following lemma provides those cocycles which are at least linear in $\psi_1$ or $\chi_1$ and in $\pb1$ or $\cb1$. 

\begin{lemma}\label{lem4D3}\quad \\
(i) Any cocycle in $\Hg(\fb)$ which is at least linear in $\psi_1$ or $\chi_1$ and in $\pb1$ or $\cb1$ is equivalent to $\tc1\chi_1\cb1-\tc3\chi_1\pb1+\tc2\psi_1\pb1-\tc4\psi_1\cb1$ times a complex number:
\begin{align}
  &\fb\om=0,\quad \om\in\Omg, \quad \om|_{\psi_1=\chi_1=0}=0,\quad \om|_{\pb1=\cb1=0}=0\quad\Then\notag\\
  &\om\sim(\tc1\chi_1\cb1-\tc3\chi_1\pb1+\tc2\psi_1\pb1-\tc4\psi_1\cb1)b,\ b\in\mathbb{C}.
 \label{4D63}
\end{align}
(ii) The cocycle $\tc1\chi_1\cb1-\tc3\chi_1\pb1+\tc2\psi_1\pb1-\tc4\psi_1\cb1$ is nontrivial in $\Hg(\fb)$:
\begin{align}
\tc1\chi_1\cb1-\tc3\chi_1\pb1+\tc2\psi_1\pb1-\tc4\psi_1\cb1\not\sim 0.
 \label{4D64}
\end{align}
\end{lemma}

{\bf Proof:} 
We expand $\om\in\Omg$ in $\psi_1$ according to
\begin{align}
  \om=\sum_{m=0}^{\ol{m}}(\psi_1)^m\om_m(\chi_1,\pb1,\cb1,\tc1,\dots,\tc4).
	\label{4D49} 
\end{align}
$\om|_{\psi_1=\chi_1=0}=0$ and $\om|_{\pb1=\cb1=0}=0$ imply that in this expansion the $\om_m$ are polynomials in $\chi_1$,$\pb1$,$\cb1$,$\tc1$,\dots,$\tc4$ that are at least linear in $\pb1$ or $\cb1$ and that $\om_0$ is at least linear in $\chi_1$:
\begin{align}
  &m>0:\ \om_m=\pb1\om_{m,1}(\chi_1,\pb1,\cb1,\tc1,\dots,\tc4)+
  \cb1\om_{m,2}(\chi_1,\cb1,\tc1,\dots,\tc4);
	\label{4D50} \\
  &\om_0=\chi_1[\pb1\om_{0,1}(\chi_1,\pb1,\cb1,\tc1,\dots,\tc4)+
  \cb1\om_{0,2}(\chi_1,\cb1,\tc1,\dots,\tc4)]
  \label{4D50a}
\end{align}
where the arguments of $\om_{m,1}$ and $\om_{m,2}$ indicate that $\om_{m,1}$ may depend polynomially on all variables $\chi_1,\pb1,\cb1,\tc1,\dots,\tc4$ whereas $\om_{m,2}$ is a polynomial only in $\chi_1,\cb1,\tc1,\dots,\tc4$ but does not depend on $\pb1$.
Notice that the expansion \eqref{4D49} is always finite since $\Omg$ is a space of polynomials in the ghost variables; hence, there is always a term with some highest degree $\ol{m}$ in $\psi_1$. Using the decomposition \eqref{4D10} of $\fb$, one infers that $\fb\om$ contains at most one term of degree $\ol{m}+1$ in $\psi_1$ given by $(\psi_1)^{\ol{m}+1}D_1 \om_{\ol{m}}$. Hence, $\fb\om=0$ requires in particular that this term vanishes,
\begin{align}
  \fb\om=0\quad\Then\quad
  D_1 \om_{\ol{m}}=0.
	\label{4D51} 
\end{align}
$D_1 \om_{\ol{m}}=0$ is treated by the "basic lemma" given in \cite{Brandt:1989rd} as follows. We introduce the antiderivation
\begin{align}
  r=\tc1\dds{\pb1}+\tc3\dds{\cb1}\, .
	\label{4D13} 
\end{align}
The anticommutator of $r$ and $D_1$ is
\begin{align}
\acom{r}{D_1}=L\, ,\quad L=N_{\pb1}+N_{\cb1}+N_{\tc1}+N_{\tc3}
	\label{4D14} 
\end{align}
with $N_{\pb1}$, $N_{\cb1}$, $N_{\tc1}$, $N_{\tc3}$ defined analogously to \eqref{4D15}.
$\om_{\ol{m}}$ is decomposed into eigenfunctions of $L$. We denote the corresponding eigenvalues by $\lambda$; these eigenvalues are positive integers since all terms in $\om_{\ol{m}}$ are at least linear in $\pb1$ or $\cb1$ owing to \eqref{4D50}, \eqref{4D50a}:
\begin{align}
  \om_{\ol{m}}=\sum_{\lambda\geq 1}\om_{\ol{m},\lambda}\, ,\quad
  L\,\om_{\ol{m},\lambda}=\lambda\,\om_{\ol{m},\lambda}\, ,\quad \lambda\in\mathbb{N}.
	\label{4D19} 
\end{align}
Owing to $\com{L}{D_1}=0$ (which follows from $L=\acom{r}{D_1}$), $D_1\om_{\ol{m}}=0$ implies $D_1\om_{\ol{m},\lambda}=0$ for all $\lambda$;
$L=\acom{D_1}{r}$ then implies that all $\om_{\ol{m},\lambda}$ are $D_1$-exact:
\begin{align}
  \forall\lambda:\quad D_1\om_{\ol{m},\lambda}=0\ \Then\ 
  \lambda\,\om_{\ol{m},\lambda}=\acom{D_1}{r}\,\om_{\ol{m},\lambda}
  =D_1r\,\om_{\ol{m},\lambda}
	\label{4D53} 
\end{align}
Hence, $\om_{\ol{m}}$ is $D_1$-exact:
\begin{align}
  \om_{\ol{m}}=D_1\sum_{\lambda\geq 1}\frac1{\lambda}\,r\,\om_{\ol{m},\lambda}\,.
	\label{4D54} 
\end{align}
This implies that we can remove the term $\om_{\ol{m}}$ of highest degree $\ol{m}$ from $\om$ by subtracting an $\fb$-coboundary, if $\ol{m}>0$:
\begin{align}
  \ol{m}>0:\ \om'&:=\om-\fb\left((\psi_1)^{\ol{m}-1}\sum_{\lambda\geq 1}
  \frac1{\lambda}\,r\,\om_{\ol{m},\lambda}\right)
  \notag\\
  &=\sum_{m=0}^{\ol{m}-1}(\psi_1)^m\om^\prime_m(\chi_1,\pb1,\cb1,\tc1,\dots,\tc4)
	\label{4D55} 
\end{align}
where 
\begin{align}
  &\om^\prime_{\ol{m}-1}=
  \om_{\ol{m}-1}-\chi_1D_2\sum_{\lambda\geq 1}\frac1{\lambda}\,r\,\om_{\ol{m},\lambda},
  \notag\\
  &m\leq\ol{m}-2:\ 
  \om^\prime_{m}=
  \om_{m}.
	\label{4D56} 
\end{align}
It should be noted that equations \eqref{4D50}, \eqref{4D50a} also apply to $\om^\prime_{\ol{m}-1}$ because $D_2$ consists of contributions that are linear in $\pb1$ or $\cb1$, cf. \eqref{4D11}.
Repeating the above procedure for $\om'$ and continuing it, one removes successively all terms depending on $\psi_1$ by subtracting $\fb$-coboundaries. This shows that $\om$ is $\fb$-exact except (possibly) for a contribution $\om^\prime_0(\chi_1,\pb1,\cb1,\tc1,\dots,\tc4)$,
\begin{align}
  \om\sim\om^\prime_0(\chi_1,\pb1,\cb1,\tc1,\dots,\tc4).
 	\label{4D57} 
\end{align}
We now expand $\om^\prime_0$ in $\chi_1$; the coefficent functions of this expansion are in $\Om^-$ defined in \eqref{4D12}:
\begin{align}
  \om^\prime_0(\chi_1,\pb1,\cb1,\tc1,\dots,\tc4)=\sum_{k\geq 1}(\chi_1)^k\7\om_k,\quad\7\om_k\in\Om^-.
 	\label{4D58} 
\end{align}
This yields
\[
\fb\om^\prime_0=\sum_{k\geq 1}(\psi_1(\chi_1)^kD_1\7\om_k+(\chi_1)^{k+1}D_2\7\om_k)
\]
and thus
\begin{align}
  \fb\om^\prime_0=0\ \LRA\ \forall k:\ D_1\7\om_k=0\ \wedge\ D_2\7\om_k=0\ 
  \LRA\ \forall k:\ \fb\7\om_k=0.
 	\label{4D59} 
\end{align}
Using now the result \eqref{4D46} and that $\om^\prime_0$ takes the form \eqref{4D50a}, we obtain
\begin{align}
  \7\om_k=&\,\pb1 a_{k}(\pb1,\cb1)+\cb1 b_k(\cb1)
  +(\tc2\pb1-\tc4\cb1)c_k(\pb1,\cb1)\notag\\
  &+(\tc1\cb1-\tc3\pb1)d_k(\pb1,\cb1)
  +(\tc1\cb1-\tc3\pb1)(\tc2\pb1-\tc4\cb1)e_k(\pb1,\cb1)
  \label{4D60}
\end{align}
for some polynomials $a_k,c_k,d_k,e_k$ in $\pb1$ and $\cb1$ and some polynomials $b_k$ in $\cb1$.
Since in $\om^\prime_0$ the various terms in \eqref{4D60} are multiplied by at least one power of $\chi_1$, cf. \eqref{4D58}, all of them provide $\fb$-exact contributions to $\om^\prime_0$ except for $d_1(\pb1,\cb1)$ because one has:
\begin{align}
  &\chi_1\pb1=\fb\tc4,\quad \chi_1\cb1=\fb\tc2,\quad
  \chi_1(\tc2\pb1-\tc4\cb1)=\fb(\tc4\tc2),
  \notag\\
  &(\chi_1)^2(\tc1\cb1-\tc3\pb1)=\fb(\tc3\tc4\chi_1-\tc1\tc2\chi_1-\tc2\tc4\psi_1),
  \notag\\
  &\chi_1(\tc1\cb1-\tc3\pb1)(\tc2\pb1-\tc4\cb1)=\fb[(\tc1\cb1-\tc3\pb1)\tc2\tc4].
 \label{4D61} 
\end{align}
The contributions to $d_1(\pb1,\cb1)$ which are at least linear in $\pb1$ or $\cb1$ also provide only $\fb$-exact contributions to $\om^\prime_0$ owing to:
\begin{align}
  &\chi_1\pb1(\tc1\cb1-\tc3\pb1)=\fb[\tc4(\tc1\cb1-\tc3\pb1)],
  \notag\\
  &\chi_1\cb1(\tc1\cb1-\tc3\pb1)=\fb[\tc2(\tc1\cb1-\tc3\pb1)].
 \label{4D61a} 
\end{align}
The only part of $d_1(\pb1,\cb1)$ which provides a possibly nontrivial contribution to $\om^\prime_0$ is thus the part which does not depend on $\pb1$ and $\cb1$ at all. We denote this part by $2b\in\mathbb{C}$ and write the corresponding contribution to $\om^\prime_0$ as:
\begin{align}
2b\,\chi_1(\tc1\cb1-\tc3\pb1)=&\,b\,(\tc1\chi_1\cb1-\tc3\chi_1\pb1+\tc2\psi_1\pb1-\tc4\psi_1\cb1)\notag\\
 &+b\,(\tc1\chi_1\cb1-\tc3\chi_1\pb1-\tc2\psi_1\pb1+\tc4\psi_1\cb1)
  \notag\\
  =&\,b\,(\tc1\chi_1\cb1-\tc3\chi_1\pb1+\tc2\psi_1\pb1-\tc4\psi_1\cb1)
  \notag\\
  &+b\,\fb(\tc3\tc4-\tc1\tc2).
   \label{4D61b}
  \end{align}
We conclude
\begin{align}
  \om^\prime_0(\chi_1,\pb1,\cb1,\tc1,\dots,\tc4)
  \sim(\tc1\chi_1\cb1-\tc3\chi_1\pb1+\tc2\psi_1\pb1-\tc4\psi_1\cb1)\,b.
 \label{4D62} 
\end{align}
Together with \eqref{4D57} this yields part (i) of lemma \ref{lem4D3}.

To prove that the cocycle $\tc1\chi_1\cb1-\tc3\chi_1\pb1+\tc2\psi_1\pb1-\tc4\psi_1\cb1$ is no coboundary in $\Hg(\fb)$, we use
arguments as in the paragraph preceding lemma 3.1 in \papII: in order to be a coboundary, $\tc1\chi_1\cb1-\tc3\chi_1\pb1+\tc2\psi_1\pb1-\tc4\psi_1\cb1$ would have to be of the form $\fb(d_{ab}c^a c^b)$ for some $d_{ab}\in\mathbb{C}$ but no such $d_{ab}$ exist. The non-existence of $d_{ab}$ can be inferred without any computation from the fact that $\tc1\chi_1\cb1-\tc3\chi_1\pb1+\tc2\psi_1\pb1-\tc4\psi_1\cb1$ is actually an \so{t,4-t}-invariant ghost polynomial, cf. section \ref{4D.2}, and therefore, owing to the \so{t,4-t}-invariance of $\fb$, $d_{ab}c^a c^b$ would have to be \so{t,4-t}-invariant too; however, there is no nonvanishing \so{t,4-t}-invariant bilinear polynomial in the translation ghosts (the only candidate bilinear polynomial would be proportional to $\eta_{ab}c^ac^b$ but this vanishes as the translation ghosts anticommute). This yields part (ii) of lemma \ref{lem4D3} and completes the proof of the lemma. \QED

{\bf Comment:} Equations \eqref{4D61b} show that the cocycle $\tc1\chi_1\cb1-\tc3\chi_1\pb1+\tc2\psi_1\pb1-\tc4\psi_1\cb1$ is equivalent to the seemingly simpler cocycle $2\chi_1(\tc1\cb1-\tc3\pb1)$. Nevertheless we prefer to use the cocycle $\tc1\chi_1\cb1-\tc3\chi_1\pb1+\tc2\psi_1\pb1-\tc4\psi_1\cb1$ owing to its \so{t,4-t}-invariance.

Lemmas \ref{lem4D1}, \ref{lem4D2} and \ref{lem4D3} provide the cohomology $\Hg(\fb)$ in the spinor representations \eqref{4D1} because the various nontrivial cocycles in these lemmas cannot combine to coboundaries. The latter statement holds because these cocycles have different degrees in $\psi_1$ and $\chi_1$ or $\pb1$ and $\cb1$ respectively, while $\fb$ increments both of these degrees by one unit. We thus conclude:

\begin{lemma}[$\Hg(\fb)$ for $\Ns=1$]\label{lem4D4}\quad \\
(i) In the spinor representations \eqref{4D1} any cocycle $\om\in\Omg$ is equivalent to a linear combination of a polynomial in $\pb1$, $\cb1$, $\tc2\pb1-\tc4\cb1$ and $\tc1\cb1-\tc3\pb1$, of a polynomial in $\psi_1$, $\chi_1$, $\tc2\psi_1-\tc3\chi_1$ and $\tc1\chi_1-\tc4\psi_1$, and of $\tc1\chi_1\cb1-\tc3\chi_1\pb1+\tc2\psi_1\pb1-\tc4\psi_1\cb1$:
\begin{align}
  \om\in\Omg:\ \fb\om=0\ \LRA\ 
  \om\sim&\,p_1(\pb1,\cb1)+q_1(\psi_1,\chi_1)\notag\\
  &+(\tc2\pb1-\tc4\cb1)p_2(\pb1,\cb1)\notag\\
  &+(\tc1\cb1-\tc3\pb1)p_3(\pb1,\cb1)\notag\\
  &+(\tc1\cb1-\tc3\pb1)(\tc2\pb1-\tc4\cb1)p_4(\pb1,\cb1)\notag\\
  &+(\tc2\psi_1-\tc3\chi_1)q_2(\psi_1,\chi_1)\notag\\
  &+(\tc1\chi_1-\tc4\psi_1)q_3(\psi_1,\chi_1)\notag\\
  &+(\tc1\chi_1-\tc4\psi_1)(\tc2\psi_1-\tc3\chi_1)q_4(\psi_1,\chi_1)\notag\\
  &+(\tc1\chi_1\cb1-\tc3\chi_1\pb1+\tc2\psi_1\pb1-\tc4\psi_1\cb1)b
  \label{4D65}
\end{align}
with polynomials $p_1(\pb1,\cb1)$,\dots, $p_4(\pb1,\cb1)$ in $\pb1$ and $\cb1$, polynomials $q_1(\psi_1,\chi_1)$,\dots, $q_4(\psi_1,\chi_1)$ in $\psi_1$ and $\chi_1$, and $b\in\mathbb{C}$.\\
(ii) A linear combination of a polynomial in $\pb1$, $\cb1$, $\tc2\pb1-\tc4\cb1$ and $\tc1\cb1-\tc3\pb1$, of a polynomial in $\psi_1$, $\chi_1$, $\tc2\psi_1-\tc3\chi_1$ and $\tc1\chi_1-\tc4\psi_1$, and of  $\tc1\chi_1\cb1-\tc3\chi_1\pb1+\tc2\psi_1\pb1-\tc4\psi_1\cb1$ is exact in $\Hg(\fb)$ if and only if it vanishes:
\begin{align}
  0\sim
  &\,p_1(\pb1,\cb1)+q_1(\psi_1,\chi_1)\notag\\
  &+(\tc2\pb1-\tc4\cb1)p_2(\pb1,\cb1)\notag\\
  &+(\tc1\cb1-\tc3\pb1)p_3(\pb1,\cb1)\notag\\
  &+(\tc1\cb1-\tc3\pb1)(\tc2\pb1-\tc4\cb1)p_4(\pb1,\cb1)\notag\\
  &+(\tc2\psi_1-\tc3\chi_1)q_2(\psi_1,\chi_1)\notag\\
  &+(\tc1\chi_1-\tc4\psi_1)q_3(\psi_1,\chi_1)\notag\\
  &+(\tc1\chi_1-\tc4\psi_1)(\tc2\psi_1-\tc3\chi_1)q_4(\psi_1,\chi_1)\notag\\
  &+(\tc1\chi_1\cb1-\tc3\chi_1\pb1+\tc2\psi_1\pb1-\tc4\psi_1\cb1)b  
  \notag\\
  \LRA\quad &
  p_1(\pb1,\cb1)+q_1(\psi_1,\chi_1)=0\ \wedge\ b=0\ \wedge\notag\\
   &\forall i\in\{2,3,4\}:\ p_i(\pb1,\cb1)=0=q_i(\psi_1,\chi_1).
  \label{4D66}
\end{align}
\end{lemma}
{\bf Comment:} Notice that in \eqref{4D66} the condition $p_1(\pb1,\cb1)+q_1(\psi_1,\chi_1)=0$ imposes that $p_1(\pb1,\cb1)$ and $q_1(\psi_1,\chi_1)$ do not depend on supersymmetry ghosts at all,
\begin{align}
  p_1(\pb1,\cb1)+q_1(\psi_1,\chi_1)=0\quad \LRA\quad p_1,q_1\in\mathbb{C}\ \wedge\ p_1=-q_1.
  \label{4D66a}
\end{align}

\subsubsection{\texorpdfstring{$\Hg(\fb)$ for $\Ns=2$}{H(gh) for N=2}}\label{4D.4}

We shall first prove the following lemma:

\begin{lemma}\label{lem4D6}
The general solution of the cocycle condition in $\Hg(\fb)$ for $\Ns=2$ in the spinor representations \eqref{4D1} is:
\begin{align}
 \fb\om=0\ \LRA\ \om\sim&\,p(\pb1,\cb1,\xi_2)
  +q(\psi_1,\chi_1,\xi_2)\notag\\
  &+(-\tc2\pb1\psi_2+\tc4\cb1\psi_2-\tc1\cb1\chi_2+\tc3\pb1\chi_2)h(\pb1,\cb1,\xi_2)\notag\\
  &+(-\tc2\psi_1\pb2+\tc3\chi_1\pb2-\tc1\chi_1\cb2+\tc4\psi_1\cb2)g(\psi_1,\chi_1,\xi_2)\notag\\
  &+(\tc1\chi_1\cb1-\tc3\chi_1\pb1+\tc2\psi_1\pb1-\tc4\psi_1\cb1\notag\\
  &\phantom{+(}-\tc1\chi_2\cb2+\tc3\chi_2\pb2-\tc2\psi_2\pb2+\tc4\psi_2\cb2)b(\xi_2)
 \label{4D86} 
\end{align}
with arbitrary polynomials $p(\pb1,\cb1,\xi_2)$, $h(\pb1,\cb1,\xi_2)$ in $\pb1$, $\cb1$, $\psi_2$, $\chi_2$, $\pb{2}$, $\cb{2}$, arbitrary polynomials $q(\psi_1,\chi_1,\xi_2)$, $g(\psi_1,\chi_1,\xi_2)$ in $\psi_1$, $\chi_1$, $\psi_2$, $\chi_2$, $\pb{2}$, $\cb{2}$, and an arbitrary polynomial $b(\xi_2)$ in $\psi_2$, $\chi_2$, $\pb{2}$, $\cb{2}$. 
\end{lemma}

{\bf Proof:}
We split the coboundary operator $\fb$ according to
\begin{align}
 \fb=\fbb1+\fbb2
 \label{4D63a} 
\end{align}
into a first operator $\fbb1$ which increments the degree in the supersymmetry ghosts $\xi_1^\ua$ ($\xi_1$-degree) by two units and a second operator $\fbb2$ which increments the degree in the supersymmetry ghosts $\xi_2^\ua$ ($\xi_2$-degree) by two units,
\begin{align}
 \fbb1=\psi_1\pb{1}\,\ddc1+\chi_1\cb{1}\,\ddc2 +\psi_1\cb{1}\,\ddc3+\chi_1\pb{1}\,\ddc4\, ,\notag\\
 \fbb2=\psi_2\pb{2}\,\ddc1+\chi_2\cb{2}\,\ddc2 +\psi_2\cb{2}\,\ddc3+\chi_2\pb{2}\,\ddc4\, .
 \label{4D63b} 
\end{align}
We denote by $N_{\xi_1}$ and $N_{\xi_2}$ the counting operators which measure the $\xi_1$-degree and $\xi_2$-degree respectively: 
\begin{align}
 N_{\xi_1}&=\xi_1^{\ua}\,\dds{\xi_1^\ua}=
 \psi_1\,\dds{\psi_1}+\chi_1\,\dds{\chi_1}+\pb1\,\dds{\pb1}+\cb1\,\dds{\cb1}\,,\notag\\
 N_{\xi_2}&=\xi_2^{\ua}\,\dds{\xi_2^\ua}=
 \psi_2\,\dds{\psi_2}+\chi_2\,\dds{\chi_2}+\pb2\,\dds{\pb2}+\cb2\,\dds{\cb2}\,.
 \label{4D63c} 
\end{align}
The first operator $\fbb1$ and the second operator $\fbb2$ and these counting operators fulfill the algebra
\begin{align}
 &(\fbb1)^2=\acom{\fbb1}{\fbb2}=(\fbb2)^2=0,\notag\\
 &\com{N_{\xi_1}}{\fbb1}=2\fbb1\, ,\quad 
 \com{N_{\xi_1}}{\fbb2}=0\, , \notag\\
 &\com{N_{\xi_2}}{\fbb1}=0\, ,\quad
 \com{N_{\xi_2}}{\fbb2}=2\fbb2\, . 
 \label{4D64a} 
\end{align}
In order to determine $\Hg(\fb)$ for $\Ns=2$ we decompose the elements $\om\in\Omg$ into eigenfunctions $\om_m$ with $\xi_1$-degree $m$:
\begin{align}
 \om=\sum_{m=0}^{\ol{m}}\om_m\,,\quad N_{\xi_1}\om_m=m\,\om_m\,.
 \label{4D67a} 
\end{align}
We shall now analyse the cocycle condition $\fb\om=0$ in $\Hg(\fb)$ by decomposing it according to the $\xi_1$-degree:
\begin{align}
 \fb\om=0\ \LRA\ \left\{
 \begin{array}{llc}
 \fbb1\om_{\ol{m}}=0,& \fbb1\om_{\ol{m}-2}+\fbb2\om_{\ol{m}}=0,& \dots \\
 \fbb1\om_{\ol{m}-1}=0,& \fbb1\om_{\ol{m}-3}+\fbb2\om_{\ol{m}-1}=0,& \dots
 \end{array}
 \right.\label{4D68a}
\end{align}
where the equations in the first line contain the $\om_{\ol{m}-2k}$ while the equations in the second line contain the  $\om_{\ol{m}-2k-1}$ ($k=0,1,\dots$). The equations of the two lines are independent and analogous to each other. Hence, it suffices to discuss the equations in the first line.

The cohomologies of $\fbb1$ and $\fbb2$ are known from lemma \ref{lem4D4}. Indeed, $\fbb1$ acts on polynomials in the $\tc{a}$, $\psi_i$, $\chi_i$, $\pb{i}$, $\cb{i}$ (with $a=1,\dots,4$ and $i=1,2$) exactly as $\fb$ in the case $\Ns=1$ on polynomials in the $\tc{a}$, $\psi_1$, $\chi_1$, $\pb{1}$, $\cb{1}$ with $\psi_2$, $\chi_2$, $\pb{2}$, $\cb{2}$ treated as ordinary complex numbers. The cohomology of $\fbb2$ is obtained from the cohomology of $\fbb1$ by interchanging the roles of $\psi_1$, $\chi_1$, $\pb{1}$, $\cb{1}$ and $\psi_2$, $\chi_2$, $\pb{2}$, $\cb{2}$.

Using the result \eqref{4D65} of lemma \ref{lem4D4} we infer from equation $\fbb1\om_{\ol{m}}=0$ in \eqref{4D68a} that
\begin{align}
 \om_{\ol{m}}=&\,\fbb1\eta_{\ol{m}-2}+p_{1,\ol{m}}(\pb1,\cb1,\xi_2)
 +q_{1,\ol{m}}(\psi_1,\chi_1,\xi_2)\notag\\
  &+(-\tc2\pb1+\tc4\cb1)p_{2,\ol{m}-1}(\pb1,\cb1,\xi_2)\notag\\
  &+(\tc1\cb1-\tc3\pb1)p_{3,\ol{m}-1}(\pb1,\cb1,\xi_2)\notag\\
  &+(\tc1\cb1-\tc3\pb1)(-\tc2\pb1+\tc4\cb1)p_{4,\ol{m}-2}(\pb1,\cb1,\xi_2)\notag\\
  &+(-\tc2\psi_1+\tc3\chi_1)q_{2,\ol{m}-1}(\psi_1,\chi_1,\xi_2)\notag\\
  &+(-\tc1\chi_1+\tc4\psi_1)q_{3,\ol{m}-1}(\psi_1,\chi_1,\xi_2)\notag\\
  &+(-\tc1\chi_1+\tc4\psi_1)(-\tc2\psi_1+\tc3\chi_1)q_{4,\ol{m}-2}(\psi_1,\chi_1,\xi_2)\notag\\
  &+\delta^2_\ol{m}\,(\tc1\chi_1\cb1-\tc3\chi_1\pb1+\tc2\psi_1\pb1-\tc4\psi_1\cb1)b(\xi_2)
 \label{4D69} 
\end{align}
for some polynomials $\eta_{\ol{m}-2}$, $p_{1,\ol{m}}(\pb1,\cb1,\xi_2)$, \dots, $b(\xi_2)$
where the subscripts $\ol{m}$, $\ol{m}-1$, $\ol{m}-2$ indicate the $\xi_1$-degree and the Kronecker symbol $\delta^2_\ol{m}$ in front of the term in the last line indicates that this term contributes only in the case $\ol{m}=2$.

We proceed to the equation $\fbb1\om_{\ol{m}-2}+\fbb2\om_{\ol{m}}=0$ in \eqref{4D68a} and use there the result \eqref{4D69} for $\om_{\ol{m}}$. By a straightforward computation this yields:
\begin{align}
 0=&\,\fbb1(\om_{\ol{m}-2}-\fbb2\eta_{\ol{m}-2})\notag\\
   &+(\pb2\cb1-\cb2\pb1)(\chi_2p_{2,\ol{m}-1}(\pb1,\cb1,\xi_2)+\psi_2p_{3,\ol{m}-1}(\pb1,\cb1,\xi_2))\notag\\
   &+(\psi_2\chi_1-\chi_2\psi_1)(\cb2q_{2,\ol{m}-1}(\psi_1,\chi_1,\xi_2)-\pb2q_{3,\ol{m}-1}(\psi_1,\chi_1,\xi_2))
   \notag\\
   &+(-\tc2\pb1+\tc4\cb1)(\pb2\cb1-\cb2\pb1)\psi_2p_{4,\ol{m}-2}(\pb1,\cb1,\xi_2)\notag\\
   &-(\tc1\cb1-\tc3\pb1)(\pb2\cb1-\cb2\pb1)\chi_2p_{4,\ol{m}-2}(\pb1,\cb1,\xi_2)\notag\\
   &-(-\tc2\psi_1+\tc3\chi_1)(\psi_2\chi_1-\chi_2\psi_1)\pb2q_{4,\ol{m}-2}(\psi_1,\chi_1,\xi_2)\notag\\
   &-(-\tc1\chi_1+\tc4\psi_1)(\psi_2\chi_1-\chi_2\psi_1)\cb2q_{4,\ol{m}-2}(\psi_1,\chi_1,\xi_2)\notag\\
   &+\delta^2_\ol{m}\,\fbb1(\tc1\chi_2\cb2-\tc3\chi_2\pb2+\tc2\psi_2\pb2-\tc4\psi_2\cb2)b(\xi_2).
 \label{4D78} 
\end{align}
The terms in the first line and in the last line of equation \eqref{4D78} are $\fbb1$-exact. Hence, the sum of the other terms is $\fbb1$-exact. Using the result \eqref{4D66} of lemma \ref{lem4D4}, we infer:
\begin{align}
 &\chi_2p_{2,\ol{m}-1}(\pb1,\cb1,\xi_2)+\psi_2p_{3,\ol{m}-1}(\pb1,\cb1,\xi_2)=0,\label{4D79}\\
 &\cb2q_{2,\ol{m}-1}(\psi_1,\chi_1,\xi_2)-\pb2q_{3,\ol{m}-1}(\psi_1,\chi_1,\xi_2)=0,\label{4D80}\\
 &p_{4,\ol{m}-2}(\pb1,\cb1,\xi_2)=0,\quad q_{4,\ol{m}-2}(\psi_1,\chi_1,\xi_2)=0.\label{4D81}
\end{align}
Equations \eqref{4D79} and \eqref{4D80} imply:
\begin{align}
 &p_{2,\ol{m}-1}(\pb1,\cb1,\xi_2)=\psi_2h_{\ol{m}-1}(\pb1,\cb1,\xi_2),\notag\\
 &p_{3,\ol{m}-1}(\pb1,\cb1,\xi_2)=-\chi_2h_{\ol{m}-1}(\pb1,\cb1,\xi_2),\notag\\
 &q_{2,\ol{m}-1}(\psi_1,\chi_1,\xi_2)=\pb2g_{\ol{m}-1}(\psi_1,\chi_1,\xi_2),\notag\\
 &q_{3,\ol{m}-1}(\psi_1,\chi_1,\xi_2)=\cb2g_{\ol{m}-1}(\psi_1,\chi_1,\xi_2)\label{4D83}
\end{align}
for some polynomials $h_{\ol{m}-1}(\pb1,\cb1,\xi_2)$ and $g_{\ol{m}-1}(\psi_1,\chi_1,\xi_2)$. Using the results \eqref{4D81} and \eqref{4D83} in equation \eqref{4D69}, the latter gives:
\begin{align}
 \om_{\ol{m}}=&\,\fbb1\eta_{\ol{m}-2}+p_{1,\ol{m}}(\pb1,\cb1,\xi_2)
  +q_{1,\ol{m}}(\psi_1,\chi_1,\xi_2)\notag\\
  &+(-\tc2\pb1\psi_2+\tc4\cb1\psi_2-\tc1\cb1\chi_2+\tc3\pb1\chi_2)h_{\ol{m}-1}(\pb1,\cb1,\xi_2)\notag\\
  &+(-\tc2\psi_1\pb2+\tc3\chi_1\pb2-\tc1\chi_1\cb2+\tc4\psi_1\cb2)g_{\ol{m}-1}(\psi_1,\chi_1,\xi_2)\notag\\
  &+\delta^2_\ol{m}\,(\tc1\chi_1\cb1-\tc3\chi_1\pb1+\tc2\psi_1\pb1-\tc4\psi_1\cb1)b(\xi_2).
 \label{4D84} 
\end{align}

Using the results \eqref{4D79} to \eqref{4D83} in equation \eqref{4D78}, the latter becomes
\begin{align}
 0=\fbb1[\om_{\ol{m}-2}-\fbb2\eta_{\ol{m}-2}
   +\delta^2_\ol{m}\,(\tc1\chi_2\cb2-\tc3\chi_2\pb2+\tc2\psi_2\pb2-\tc4\psi_2\cb2)b(\xi_2)].
 \label{4D85} 
\end{align}
This is an equation like $\fbb1\om_{\ol{m}}=0$ in \eqref{4D68a}, with $\om_{\ol{m}-2}-\fbb2\eta_{\ol{m}-2}
+\delta^2_\ol{m}\,(\tc1\chi_2\cb2-\tc3\chi_2\pb2+\tc2\psi_2\pb2-\tc4\psi_2\cb2)b(\xi_2)$ in place of $\om_{\ol{m}}$. One can analyse this equation as $\fbb1\om_{\ol{m}}=0$ above and obtains that $\om_{\ol{m}-2}$ is given by terms as in equation \eqref{4D84} plus the $\fbb2$-coboundary $\fbb2\eta_{\ol{m}-2}$ and the term $(\tc1\chi_2\cb2-\tc3\chi_2\pb2+\tc2\psi_2\pb2-\tc4\psi_2\cb2)b(\xi_2)$ in the case $\ol{m}=2$.
Proceeding analogously to terms of lower $\xi_1$-degree, one obtains lemma \ref{lem4D6}. \QED

To completely characterize $\Hg(\fb)$ for $\Ns=2$ in the spinor representation \eqref{4D1}, we still have to determine those cocycles occurring in \eqref{4D86} that are coboundaries in $\Hg(\fb)$. In other words: we still have to determine those ghost polynomials $p(\pb1,\cb1,\xi_2)$, $q(\psi_1,\chi_1,\xi_2)$, $h(\pb1,\cb1,\xi_2)$, $g(\psi_1,\chi_1,\xi_2)$ and $b(\xi_2)$ for which the cocycles given in \eqref{4D86} combine to a coboundary in $\Hg(\fb)$. The solution to this problem is the following lemma \ref{lem4D7} which together with lemma \ref{lem4D6} provides an exhaustive characterization of $\Hg(\fb)$ for $\Ns=2$ in the spinor representations \eqref{4D1}.

\begin{lemma}[Coboundaries in lemma \ref{lem4D6}]\label{lem4D7}
\begin{align}
 &0\sim p(\pb1,\cb1,\xi_2)
  +q(\psi_1,\chi_1,\xi_2)\notag\\
  &\phantom{0\sim}+(-\tc2\pb1\psi_2+\tc4\cb1\psi_2-\tc1\cb1\chi_2+\tc3\pb1\chi_2)h(\pb1,\cb1,\xi_2)\notag\\
  &\phantom{0\sim}+(-\tc2\psi_1\pb2+\tc3\chi_1\pb2-\tc1\chi_1\cb2+\tc4\psi_1\cb2)g(\psi_1,\chi_1,\xi_2)\notag\\
  &\phantom{0\sim}+(\tc1\chi_1\cb1-\tc3\chi_1\pb1+\tc2\psi_1\pb1-\tc4\psi_1\cb1\notag\\
  &\phantom{0\sim+(}-\tc1\chi_2\cb2+\tc3\chi_2\pb2-\tc2\psi_2\pb2+\tc4\psi_2\cb2)b(\xi_2)\notag\\
  \LRA\quad&
   p(\pb1,\cb1,\xi_2)+q(\psi_1,\chi_1,\xi_2)\notag\\
   &\quad=(\pb2\cb1-\cb2\pb1)(\chi_2\4p_{2}(\pb1,\cb1,\xi_2)+\psi_2\4p_{3}(\pb1,\cb1,\xi_2))\notag\\
   &\quad\phantom{=}+(\psi_2\chi_1-\chi_2\psi_1)(\cb2\4q_{2}(\psi_1,\chi_1,\xi_2)
         -\pb2\4q_{3}(\psi_1,\chi_1,\xi_2)),\notag\\
 &h(\pb1,\cb1,\xi_2)=(\pb2\cb1-\cb2\pb1)\4p_{4}(\pb1,\cb1,\xi_2),\notag\\
 &g(\psi_1,\chi_1,\xi_2)=-(\psi_2\chi_1-\chi_2\psi_1)\4q_{4}(\psi_1,\chi_1,\xi_2),\notag\\
 &b(\xi_2)=0
 \label{4D96} 
\end{align}
for polynomials $\4p_{2}(\pb1,\cb1,\xi_2)$, $\4p_{3}(\pb1,\cb1,\xi_2)$, $\4p_{4}(\pb1,\cb1,\xi_2)$ in $\pb1$, $\cb1$, $\psi_2$, $\chi_2$, $\pb{2}$, $\cb{2}$, and polynomials $\4q_{2}(\psi_1,\chi_1,\xi_2)$, 
$\4q_{3}(\psi_1,\chi_1,\xi_2)$, $\4q_{4}(\psi_1,\chi_1,\xi_2)$ in $\psi_1$, $\chi_1$, $\psi_2$, $\chi_2$, $\pb{2}$, $\cb{2}$. 
\end{lemma}

{\bf Proof:}
We study the coboundary condition
\begin{align}
  \om=\fb\eta,\ \om=&\, p(\pb1,\cb1,\xi_2)+q(\psi_1,\chi_1,\xi_2)\notag\\
  &+(-\tc2\pb1\psi_2+\tc4\cb1\psi_2-\tc1\cb1\chi_2+\tc3\pb1\chi_2)h(\pb1,\cb1,\xi_2)\notag\\
  &+(-\tc2\psi_1\pb2+\tc3\chi_1\pb2-\tc1\chi_1\cb2+\tc4\psi_1\cb2)g(\psi_1,\chi_1,\xi_2)\notag\\
  &+(\tc1\chi_1\cb1-\tc3\chi_1\pb1+\tc2\psi_1\pb1-\tc4\psi_1\cb1\notag\\
  &\phantom{+(}-\tc1\chi_2\cb2+\tc3\chi_2\pb2-\tc2\psi_2\pb2+\tc4\psi_2\cb2)b(\xi_2).
 \label{4D87} 
\end{align}
Again, we use a decomposition according to the $\xi_1$-degree:
\begin{align}
  &\om=\sum_{m=0}^{\ol{m}}\om_m\,,\quad N_{\xi_1}\om_m=m\,\om_m\,,\notag\\
  &\eta=\sum_{m=0}^{\ol{p}}\eta_m\,,\quad N_{\xi_1}\eta_m=m\,\eta_m
 \label{4D88} 
\end{align}
where $\om_{\ol{m}}$ and $\eta_{\ol{p}}$ do not vanish, respectively.  

If $\ol{p}> \ol{m}$, the coboundary condition \eqref{4D87} yields $\fbb1\eta_\ol{p}=0$ and $\fbb1\eta_{\ol{p}-2}+\fbb2\eta_{\ol{p}}=0$ at $\xi_1$-degrees $\ol{p}+2$ and $\ol{p}$, respectively. These equations imply by the same analysis as in the proof of lemma \ref{lem4D6} that $\eta_{\ol{p}}$ is of the form given in \eqref{4D84}. Since contributions to $\eta$ of that form provide cocycles in $\Hg(\fb)$, they do not contribute to the coboundary condition \eqref{4D87} and are thus irrelevant to this coboundary condition. Hence, with no loss of generality we can assume $\ol{p}\leq \ol{m}$. 

If $\ol{p}< \ol{m}-2$, the coboundary condition \eqref{4D87} yields $\om_{\ol{m}}=0$ at $\xi_1$-degree $\ol{m}$ which contradicts that $\om_{\ol{m}}$ does not vanish. 

If $\ol{p}= \ol{m}-2$ or $\ol{p}= \ol{m}-1$ the coboundary condition \eqref{4D87} yields $\om_{\ol{m}}=\fbb1\eta_{\ol{m}-2}$ at $\xi_1$-degree $\ol{m}$, i.e., $\om_{\ol{m}}$ is $\fbb1$-exact. This implies $\om_{\ol{m}}=0$ by the result \eqref{4D66} in lemma \ref{lem4D4} and also contradicts that $\om_{\ol{m}}$ does not vanish. 

Hence, with no loss of generality we can assume $\ol{p}= \ol{m}$ which yields the following decomposition of the coboundary condition \eqref{4D87}:
\begin{align}
  &\fbb1\eta_{\ol{m}}=0,\quad \fbb1\eta_{\ol{m}-1}=0,\label{4D87a}\\
  &\om_{m}=\fbb1\eta_{m-2}+\fbb2\eta_{m}\quad \mbox{for}\ m=2,\dots,\ol{m},\label{4D87b}\\
  &\om_{1}=\fbb2\eta_{1},\quad \om_{0}=\fbb2\eta_{0}.\label{4D87c}
\end{align}
From the first equation \eqref{4D87a} we infer by the same arguments
that led to equation \eqref{4D69}:
\begin{align}
 \eta_{\ol{m}}=&\,\fbb1\4\eta_{\ol{m}-2}+\4p_{1,\ol{m}}(\pb1,\cb1,\xi_2)
 +\4q_{1,\ol{m}}(\psi_1,\chi_1,\xi_2)\notag\\
  &+(-\tc2\pb1+\tc4\cb1)\4p_{2,\ol{m}-1}(\pb1,\cb1,\xi_2)\notag\\
  &+(\tc1\cb1-\tc3\pb1)\4p_{3,\ol{m}-1}(\pb1,\cb1,\xi_2)\notag\\
  &+(\tc1\cb1-\tc3\pb1)(-\tc2\pb1+\tc4\cb1)\4p_{4,\ol{m}-2}(\pb1,\cb1,\xi_2)\notag\\
  &+(-\tc2\psi_1+\tc3\chi_1)\4q_{2,\ol{m}-1}(\psi_1,\chi_1,\xi_2)\notag\\
  &+(-\tc1\chi_1+\tc4\psi_1)\4q_{3,\ol{m}-1}(\psi_1,\chi_1,\xi_2)\notag\\
  &+(-\tc1\chi_1+\tc4\psi_1)(-\tc2\psi_1+\tc3\chi_1)\4q_{4,\ol{m}-2}(\psi_1,\chi_1,\xi_2)\notag\\
  &+\delta^2_\ol{m}\,(\tc1\chi_1\cb1-\tc3\chi_1\pb1+\tc2\psi_1\pb1-\tc4\psi_1\cb1)\4b(\xi_2).
 \label{4D89} 
\end{align}

The second equation \eqref{4D87a} implies an analogous result for $\eta_{\ol{m}-1}$.

Using the result \eqref{4D89} in the equation \eqref{4D87b} for $m=\ol{m}$, we obtain
\begin{align}
 \om_{\ol{m}}=
   &\,(\pb2\cb1-\cb2\pb1)(\chi_2\4p_{2,\ol{m}-1}(\pb1,\cb1,\xi_2)+\psi_2\4p_{3,\ol{m}-1}(\pb1,\cb1,\xi_2))\notag\\
   &+(\psi_2\chi_1-\chi_2\psi_1)(\cb2\4q_{2,\ol{m}-1}(\psi_1,\chi_1,\xi_2)-\pb2\4q_{3,\ol{m}-1}(\psi_1,\chi_1,\xi_2))
   \notag\\
   &+(-\tc2\pb1+\tc4\cb1)(\pb2\cb1-\cb2\pb1)\psi_2\4p_{4,\ol{m}-2}(\pb1,\cb1,\xi_2)\notag\\
   &-(\tc1\cb1-\tc3\pb1)(\pb2\cb1-\cb2\pb1)\chi_2\4p_{4,\ol{m}-2}(\pb1,\cb1,\xi_2)\notag\\
   &-(-\tc2\psi_1+\tc3\chi_1)(\psi_2\chi_1-\chi_2\psi_1)\pb2\4q_{4,\ol{m}-2}(\psi_1,\chi_1,\xi_2)\notag\\
   &-(-\tc1\chi_1+\tc4\psi_1)(\psi_2\chi_1-\chi_2\psi_1)\cb2\4q_{4,\ol{m}-2}(\psi_1,\chi_1,\xi_2)\notag\\
   &+\delta^2_\ol{m}\,\fbb1(\tc1\chi_2\cb2-\tc3\chi_2\pb2+\tc2\psi_2\pb2-\tc4\psi_2\cb2)\4b(\xi_2)\notag\\
   &+\fbb1(\eta_{\ol{m}-2}-\fbb2\4\eta_{\ol{m}-2}).
 \label{4D90} 
\end{align}
This equation states that $\om_{\ol{m}}$ minus the terms in the first six lines on the right hand side is $\fbb1$-exact. Using the result \eqref{4D66} and the explicit form of $\om$ given in \eqref{4D87}, we infer from equation \eqref{4D90}:
\begin{align}
 p_{\ol{m}}+q_{\ol{m}}&=
   (\pb2\cb1-\cb2\pb1)(\chi_2\4p_{2,\ol{m}-1}(\pb1,\cb1,\xi_2)+\psi_2\4p_{3,\ol{m}-1}(\pb1,\cb1,\xi_2))\notag\\
   &\phantom{=}+(\psi_2\chi_1-\chi_2\psi_1)(\cb2\4q_{2,\ol{m}-1}(\psi_1,\chi_1,\xi_2)
         -\pb2\4q_{3,\ol{m}-1}(\psi_1,\chi_1,\xi_2)),
   \label{4D91}\\
 h_{\ol{m}-1}&=(\pb2\cb1-\cb2\pb1)\4p_{4,\ol{m}-2}(\pb1,\cb1,\xi_2),\label{4D92}\\
 g_{\ol{m}-1}&=-(\psi_2\chi_1-\chi_2\psi_1)\4q_{4,\ol{m}-2}(\psi_1,\chi_1,\xi_2),\label{4D93}\\
 \delta^2_\ol{m}\,b(\xi_2)&=0
 \label{4D94} 
\end{align}
where $p_{\ol{m}}$ denotes the contribution to $p(\pb1,\cb1,\xi_2)$ with $\xi_1$-degree $\ol{m}$ et cetera.

Using the results \eqref{4D91} to \eqref{4D94} in \eqref{4D90}, the latter yields furthermore
\begin{align}
 \fbb1(\eta_{\ol{m}-2}-\fbb2\4\eta_{\ol{m}-2}
 +\delta^2_\ol{m}\,(\tc1\chi_2\cb2-\tc3\chi_2\pb2+\tc2\psi_2\pb2-\tc4\psi_2\cb2)\4b(\xi_2))=0.
 \label{4D95} 
\end{align}
Equation \eqref{4D95} implies that $\eta_{\ol{m}-2}-\fbb2\4\eta_{\ol{m}-2}
 +\delta^2_\ol{m}\,(\tc1\chi_2\cb2-\tc3\chi_2\pb2+\tc2\psi_2\pb2-\tc4\psi_2\cb2)\4b(\xi_2)$ is of the same form as $\eta_{\ol{m}}$ in equation \eqref{4D89}. One can continue the analysis of equations \eqref{4D87b} and \eqref{4D87c} analogously to lower $\xi_1$-degrees. This yields results analogous to \eqref{4D91} to \eqref{4D94} for the other contributions $p_m$, $q_m$, $h_m$, $g_m$ to the polynomials $p$, $q$, $h$, $g$ in $\om$ and completes the proof of lemma \ref{lem4D7}. \QED

\subsubsection{\texorpdfstring{$\Hg(\fb)$ for $\Ns>2$}{H(gh) for N>2}}\label{4D.5}

We shall determine $\Hg(\fb)$ for $\Ns>2$ using the results for $\Ns=2$ by a strategy analogous to the strategy we have used to determine $\Hg(\fb)$ for $\Ns=2$ by means of the results for $\Ns=1$ in section \ref{4D.4} and shall first prove the following result:

\begin{lemma}\label{lem4D9}
The general solution of the cocycle condition in $\Hg(\fb)$ for $\Ns>2$ in the spinor representation \eqref{4D1} is:
\begin{align}
 \fb\om=0\ \LRA\ \om\sim&\,p(\pb1,\cb1,\xi_2,\dots,\xi_\Ns)
  +q(\psi_1,\chi_1,\xi_2,\dots,\xi_\Ns)
 \label{4D113} 
\end{align}
with an arbitrary polynomial $p(\pb1,\cb1,\xi_2,\dots,\xi_\Ns)$ in $\pb1$, $\cb1$, $\psi_2$, \dots, $\psi_\Ns$, $\chi_2$, \dots, $\chi_\Ns$, $\pb{2}$, \dots , $\pb{\Ns}$, $\cb{2}$, \dots, $\cb{\Ns}$, and an arbitrary polynomial $q(\psi_1,\chi_1,\xi_2,\dots,\xi_\Ns)$ in $\psi_1$, $\chi_1$, $\psi_2$, \dots, $\psi_\Ns$, $\chi_2$, \dots, $\chi_\Ns$, $\pb{2}$, \dots , $\pb{\Ns}$, $\cb{2}$, \dots, $\cb{\Ns}$.
\end{lemma}

{\bf Proof:}
We split
the coboundary operator $\fb$ according to
\begin{align}
 \fb=\fbb{\Ns=2}+\fbb{\Ns>2}
 \label{4D101} 
\end{align}
into a first operator $\fbb{\Ns=2}$ which acts like $\fb$ in the case $\Ns=2$, and a second operator $\fbb{\Ns>2}$ which contains the remaining terms of $\fb$: 
\begin{align}
 &\fbb{\Ns=2}=\sum_{i=1}^{2}\left(\psi_i\pb{i}\,\ddc1+\chi_i\cb{i}\,\ddc2 
 +\psi_i\cb{i}\,\ddc3+\chi_i\pb{i}\,\ddc4\right) ,\notag\\
 &\fbb{\Ns>2}=\sum_{i=3}^{\Ns}\left(\psi_i\pb{i}\,\ddc1+\chi_i\cb{i}\,\ddc2 
 +\psi_i\cb{i}\,\ddc3+\chi_i\pb{i}\,\ddc4\right) .
 \label{4D102} 
\end{align}
We denote by $N_{\Ns=2}$ the counting operator which measures the degree of homogeneity in the components of the supersymmetry ghosts $\xi_1$ and $\xi_2$, and by $N_{\Ns>2}$ the counting operator which measures the degree homogeneity in the components of the remaining supersymmetry ghosts $\xi_3$, \dots, $\xi_\Ns$:
\begin{align}
 N_{\Ns=2}=\sum_{i=1}^{2}\xi_i^{\ua}\,\dds{\xi_i^\ua}\,,\quad
 N_{\Ns>2}=\sum_{i=3}^{\Ns}\xi_i^{\ua}\,\dds{\xi_i^\ua}\,.
 \label{4D103} 
\end{align}
The first operator $\fbb{\Ns=2}$ and the second operator $\fbb{\Ns>2}$ and these counting operators fulfill an algebra analogous to \eqref{4D64a}:
\begin{align}
 &(\fbb{\Ns=2})^2=\acom{\fbb{\Ns=2}}{\fbb{\Ns>2}}=(\fbb{\Ns>2})^2=0,\notag\\
 &\com{N_{\Ns=2}}{\fbb{\Ns=2}}=2\fbb{\Ns=2}\, ,\quad 
 \com{N_{\Ns=2}}{\fbb{\Ns>2}}=0\, , \notag\\
 &\com{N_{\Ns>2}}{\fbb{\Ns=2}}=0\, ,\quad
 \com{N_{\Ns>2}}{\fbb{\Ns=2}}=2\fbb{\Ns=2}\, . 
 \label{4D104} 
\end{align}
The elements $\om\in\Omg$ are decomposed into eigenfunctions $\om_m$ of $N_{\Ns=2}$ with eigenvalue $m$,
\begin{align}
 \om=\sum_{m=0}^{\ol{m}}\om_m\,,\quad N_{\Ns=2}\om_m=m\,\om_m\,,
 \label{4D105} 
\end{align}
and the cocycle condition $\fb\om=0$ in $\Hg(\fb)$ is decomposed accordingly into
\begin{align}
 \fb\om=0\ \LRA\ \left\{
 \begin{array}{llc}
 \fbb{\Ns=2}\om_{\ol{m}}=0,& \fbb{\Ns=2}\om_{\ol{m}-2}+\fbb{\Ns>2}\om_{\ol{m}}=0,& \dots \\
 \fbb{\Ns=2}\om_{\ol{m}-1}=0,& \fbb{\Ns=2}\om_{\ol{m}-3}+\fbb{\Ns>2}\om_{\ol{m}-1}=0,& \dots
 \end{array}
 \right.
 \label{4D106} 
\end{align}
where, as in equations \eqref{4D68a}, the equations in the first and second line are analysed independently and analogously to each other. 
Using lemma \ref{lem4D6} we infer from equation $\fbb{\Ns=2}\om_{\ol{m}}=0$ in \eqref{4D106} that
\begin{align}
 \om_{\ol{m}}=&\,\fbb{\Ns=2}\eta_{\ol{m}-2}+p_{\ol{m}}(\pb1,\cb1,\xi_2,\dots,\xi_\Ns)
  +q_{\ol{m}}(\psi_1,\chi_1,\xi_2,\dots,\xi_\Ns)\notag\\
  &+(-\tc2\pb1\psi_2+\tc4\cb1\psi_2-\tc1\cb1\chi_2+\tc3\pb1\chi_2)h_{\ol{m}-2}(\pb1,\cb1,\xi_2,\dots,\xi_\Ns)\notag\\
  &+(-\tc2\psi_1\pb2+\tc3\chi_1\pb2-\tc1\chi_1\cb2+\tc4\psi_1\cb2)
  g_{\ol{m}-2}(\psi_1,\chi_1,\xi_2,\dots,\xi_\Ns)\notag\\
  &+(\tc1\chi_1\cb1-\tc3\chi_1\pb1+\tc2\psi_1\pb1-\tc4\psi_1\cb1\notag\\
  &\phantom{+(}-\tc1\chi_2\cb2+\tc3\chi_2\pb2-\tc2\psi_2\pb2+\tc4\psi_2\cb2)b_{\ol{m}-2}(\xi_2,\dots,\xi_\Ns).
 \label{4D107} 
\end{align}

Using the result \eqref{4D107} for $\om_{\ol{m}}$ in the equation $\fbb{\Ns=2}\om_{\ol{m}-2}+\fbb{\Ns>2}\om_{\ol{m}}=0$ in \eqref{4D106}, we obtain
\begin{align}
0=&\,\fbb{\Ns=2}(\om_{\ol{m}-2}-\fbb{\Ns>2}\eta_{\ol{m}-2})\notag\\
  &+h_{\ol{m}-2}(\pb1,\cb1,\xi_2,\dots,\xi_\Ns)
  \sum_{i=3}^\Ns(-\chi_i\cb{i}\pb1\psi_2
  +\chi_i\pb{i}\cb1\psi_2-\psi_i\pb{i}\cb1\chi_2\notag\\
  &+\psi_i\cb{i}\pb1\chi_2)
  +g_{\ol{m}-2}(\psi_1,\chi_1,\xi_2,\dots,\xi_\Ns)
  \sum_{i=3}^\Ns(-\chi_i\cb{i}\psi_1\pb2
  +\psi_i\cb{i}\chi_1\pb2\notag\\
  &-\psi_i\pb{i}\chi_1\cb2+\chi_i\pb{i}\psi_1\cb2)
  + b_{\ol{m}-2}(\xi_2,\dots,\xi_\Ns)
  \sum_{i=3}^\Ns(\psi_i\pb{i}(\chi_1\cb1-\chi_2\cb2)\notag\\
  &-\psi_i\cb{i}(\chi_1\pb1-\chi_2\pb2)+\chi_i\cb{i}(\psi_1\pb1-\psi_2\pb2)
  -\chi_i\pb{i}(\psi_1\cb1-\psi_2\cb2))
  \notag\\[8pt]
 =&\,\fbb{\Ns=2}(\om_{\ol{m}-2}-\fbb{\Ns>2}\eta_{\ol{m}-2})\notag\\
  &+h_{\ol{m}-2}(\pb1,\cb1,\xi_2,\dots,\xi_\Ns)
  \sum_{i=3}^\Ns(\psi_i\chi_2-\chi_i\psi_2)(\cb{i}\pb1-\pb{i}\cb1)\notag\\
  &+g_{\ol{m}-2}(\psi_1,\chi_1,\xi_2,\dots,\xi_\Ns)
  \sum_{i=3}^\Ns(\psi_i\chi_1-\chi_i\psi_1)(\cb{i}\pb2-\pb{i}\cb2)\notag\\
  &+\fbb{\Ns=2}\Big(b_{\ol{m}-2}(\xi_2,\dots,\xi_\Ns)
  \sum_{i=3}^\Ns(\psi_i\pb{i}\tc2
  -\psi_i\cb{i}\tc4+\chi_i\cb{i}\tc1-\chi_i\pb{i}\tc3)\Big)\notag\\
  &- 2b_{\ol{m}-2}(\xi_2,\dots,\xi_\Ns)
  \sum_{i=3}^\Ns(\chi_i\psi_2-\psi_i\chi_2)(\cb{i}\pb2-\pb{i}\cb2).
 \label{4D108} 
\end{align}
Using lemma \ref{lem4D7} we conclude from equation \eqref{4D108}:
\begin{align}
  &h_{\ol{m}-2}(\pb1,\cb1,\xi_2,\dots,\xi_\Ns)
  \sum_{i=3}^\Ns(\psi_i\chi_2-\chi_i\psi_2)(\cb{i}\pb1-\pb{i}\cb1)\notag\\
  &+g_{\ol{m}-2}(\psi_1,\chi_1,\xi_2,\dots,\xi_\Ns)
  \sum_{i=3}^\Ns(\psi_i\chi_1-\chi_i\psi_1)(\cb{i}\pb2-\pb{i}\cb2)\notag\\
  &- 2b_{\ol{m}-2}(\xi_2,\dots,\xi_\Ns)
  \sum_{i=3}^\Ns(\chi_i\psi_2-\psi_i\chi_2)(\cb{i}\pb2-\pb{i}\cb2)\notag\\
  &=(\pb2\cb1-\cb2\pb1)(\chi_2\4p_{2,\ol{m}-2}(\pb1,\cb1,\xi_2,\dots,\xi_\Ns)
  +\psi_2\4p_{3,\ol{m}-2}(\pb1,\cb1,\xi_2,\dots,\xi_\Ns))\notag\\
  &\phantom{=}+
  (\psi_2\chi_1-\chi_2\psi_1)(\cb2\4q_{2,\ol{m}-2}(\psi_1,\chi_1,\xi_2,\dots,\xi_\Ns)
  -\pb2\4q_{3,\ol{m}-2}(\psi_1,\chi_1,\xi_2,\dots,\xi_\Ns))
 \label{4D109a} 
\end{align}
for some polynomials  $\4p_{2,\ol{m}-2}(\pb1,\cb1,\xi_2,\dots,\xi_\Ns)$,\dots, $\4q_{3,\ol{m}-2}(\psi_1,\chi_1,\xi_2,\dots,\xi_\Ns)$. By inspecting the dependence on $\psi_1$, $\chi_1$, $\pb1$ and $\cb1$ of the various terms in equation \eqref{4D109a}, we obtain
\begin{align}
 h_{\ol{m}-2}(\pb1,\cb1,\xi_2,\dots,\xi_\Ns)=&\,
 (\pb2\cb1-\cb2\pb1)u_{\ol{m}-4}(\pb1,\cb1,\xi_2,\dots,\xi_\Ns),\notag\\
 g_{\ol{m}-2}(\psi_1,\chi_1,\xi_2,\dots,\xi_\Ns)=&\,
 (\psi_2\chi_1-\chi_2\psi_1)v_{\ol{m}-4}(\psi_1,\chi_1,\xi_2,\dots,\xi_\Ns),\notag\\
 b_{\ol{m}-2}(\xi_2,\dots,\xi_\Ns)=&\,0
 \label{4D109} 
\end{align}
for some polynomials $u_{\ol{m}-4}(\pb1,\cb1,\xi_2,\dots,\xi_\Ns)$ and $v_{\ol{m}-4}(\psi_1,\chi_1,\xi_2,\dots,\xi_\Ns)$.

Using equations \eqref{4D109} in \eqref{4D107} yields
\begin{align}
 &\om_{\ol{m}}=\fbb{\Ns=2}\eta_{\ol{m}-2}+p_{\ol{m}}(\pb1,\cb1,\xi_2,\dots,\xi_\Ns)
  +q_{\ol{m}}(\psi_1,\chi_1,\xi_2,\dots,\xi_\Ns)\notag\\
  &+(\tc4\cb1\psi_2-\tc2\pb1\psi_2+\tc3\pb1\chi_2-\tc1\cb1\chi_2)(\pb2\cb1-\cb2\pb1)
  u_{\ol{m}-4}\notag\\
 &+(\tc3\chi_1\pb2-\tc2\psi_1\pb2+\tc4\psi_1\cb2-\tc1\chi_1\cb2)
  (\psi_2\chi_1-\chi_2\psi_1)
  v_{\ol{m}-4}\notag\\
 &=\fbb{\Ns=2}\4\eta_{\ol{m}-2}+p_{\ol{m}}(\pb1,\cb1,\xi_2,\dots,\xi_\Ns)
  +q_{\ol{m}}(\psi_1,\chi_1,\xi_2,\dots,\xi_\Ns) 
 \label{4D110} 
\end{align}
where the arguments of $u_{\ol{m}-4}$ and $v_{\ol{m}-4}$ have been left out and 
\begin{align}
\4\eta_{\ol{m}-2}=&\,\eta_{\ol{m}-2}
  +(\tc2\pb1-\tc4\cb1)(\tc1\cb1-\tc3\pb1)
   u_{\ol{m}-4}\notag\\
   &+(\tc3\chi_1-\tc2\psi_1)(\tc4\psi_1-\tc1\chi_1)
   v_{\ol{m}-4}.
 \label{4D111} 
\end{align}
According to \eqref{4D110}, $\om_{\ol{m}}$ is $\fbb{\Ns=2}$-exact up to (possibly) terms $p_{\ol{m}}(\pb1,\cb1,\xi_2,\dots,\xi_\Ns)
+q_{\ol{m}}(\psi_1,\chi_1,\xi_2,\dots,\xi_\Ns)$. The latter terms are $\fb$-closed. Hence, the polynomial
\begin{align}
\om^\prime=\om-\fb\4\eta_{\ol{m}-2}-p_{\ol{m}}(\pb1,\cb1,\xi_2,\dots,\xi_\Ns)
-q_{\ol{m}}(\psi_1,\chi_1,\xi_2,\dots,\xi_\Ns)
 \label{4D112} 
\end{align}
is $\fb$-closed and its decomposition \eqref{4D105} contains only terms with $N_{\Ns=2}$-eigenvalues $m<\ol{m}$. $\om^\prime$ is then treated as $\om$ before, leading to a result analogous to \eqref{4D110} for the contribution $\om^\prime_{\ol{m}^\prime}$ with highest $N_{\Ns=2}$-eigenvalue $\ol{m}^\prime$ contained in $\om^\prime$ (where $\ol{m}^\prime<\ol{m}$). Continuing the arguments, one concludes that $\om$ is $\fb$-exact except, possibly, for terms $p(\pb1,\cb1,\xi_2,\dots,\xi_\Ns)
+q(\psi_1,\chi_1,\xi_2,\dots,\xi_\Ns)=\sum_{m=0}^{\ol{m}}(p_m(\pb1,\cb1,\xi_2,\dots,\xi_\Ns)
+q_m(\psi_1,\chi_1,\xi_2,\dots,\xi_\Ns))$. This yields lemma \ref{lem4D9}. \QED

To complete the computation of $\Hg(\fb)$ for $\Ns>2$ we still have to determine those polynomials  $p(\pb1,\cb1,\xi_2,\dots,\xi_\Ns)+q(\psi_1,\chi_1,\xi_2,\dots,\xi_\Ns)$ which are $\fb$-exact. 
The solution to this problem is the following lemma \ref{lem4D10} which together with lemma \ref{lem4D9} provides an exhaustive characterization of $\Hg(\fb)$ for $\Ns>2$ in the spinor representations \eqref{4D1}.

\begin{lemma}[Coboundaries in lemma \ref{lem4D9}]\label{lem4D10}
\begin{align}
  &p(\pb1,\cb1,\xi_2,\dots,\xi_\Ns)+q(\psi_1,\chi_1,\xi_2,\dots,\xi_\Ns)\sim 0\notag\\
  \LRA\ &
  p(\pb1,\cb1,\xi_2,\dots,\xi_\Ns)+q(\psi_1,\chi_1,\xi_2,\dots,\xi_\Ns)=\fb\7\eta,\notag\\
  &\7\eta=
  \Big[(\tc1\cb{1}\chi_1-\tc3\pb{1}\chi_1+\tc2\pb{1}\psi_1-\tc4\cb{1}\psi_1)
  \notag\\
  &\phantom{\7\eta=}-\sum_{i=2}^\Ns(\tc1\cb{i}\chi_i-\tc3\pb{i}\chi_i+\tc2\pb{i}\psi_i-\tc4\cb{i}\psi_i)\Big]
  \7b(\xi_2,\dots,\xi_\Ns)
  \notag\\
  &\phantom{\7\eta=}+(\tc4\cb1-\tc2\pb1)\4p_{2}(\pb1,\cb1,\xi_2,\dots,\xi_\Ns)\notag\\
  &\phantom{\7\eta=}+(\tc1\cb1-\tc3\pb1)\4p_{3}(\pb1,\cb1,\xi_2,\dots,\xi_\Ns)\notag\\
  &\phantom{\7\eta=}+(\tc3\chi_1-\tc2\psi_1)\4q_{2}(\psi_1,\chi_1,\xi_2,\dots,\xi_\Ns)\notag\\
  &\phantom{\7\eta=}+(\tc4\psi_1-\tc1\chi_1)\4q_{3}(\psi_1,\chi_1,\xi_2,\dots,\xi_\Ns)\notag\\
  \LRA\ &p(\pb1,\cb1,\xi_2,\dots,\xi_\Ns)+q(\psi_1,\chi_1,\xi_2,\dots,\xi_\Ns)=\notag\\
  &-\7b(\xi_2,\dots,\xi_\Ns)
  \sum_{i=2}^\Ns\sum_{j=2}^\Ns(\pb{i}\cb{j}-\cb{i}\pb{j})(\psi_i\chi_j-\chi_i\psi_j)
  \notag\\
  &+\sum_{i=2}^\Ns\Big[(\psi_i\chi_1-\chi_i\psi_1)(\cb{i}\4q_{2}(\psi_1,\chi_1,\xi_2,\dots,\xi_\Ns)
   -\pb{i}\4q_{3}(\psi_1,\chi_1,\xi_2,\dots,\xi_\Ns))
  \notag\\
  &+(\pb{i}\cb1-\cb{i}\pb1)(\chi_i\4p_{2}(\pb1,\cb1,\xi_2,\dots,\xi_\Ns)
   +\psi_i\4p_{3}(\pb1,\cb1,\xi_2,\dots,\xi_\Ns))\Big].
 \label{4D224} 
\end{align}
\end{lemma}

{\bf Proof:}
We study the coboundary condition
\begin{align}
  p(\pb1,\cb1,\xi_2,\dots,\xi_\Ns)+q(\psi_1,\chi_1,\xi_2,\dots,\xi_\Ns)=\fb\eta
 \label{4D113a} 
\end{align}
by decomposing it according to $N_{\Ns=2}$-eigenvalues, using
\begin{align}
  &p(\pb1,\cb1,\xi_2,\dots,\xi_\Ns)=\sum_{m=0}^{\ol{m}}p_m\,,\quad N_{\Ns=2}p_m=m\,p_m\,,\notag\\
  &q(\psi_1,\chi_1,\xi_2,\dots,\xi_\Ns)=\sum_{m=0}^{\ol{m}}q_m\,,\quad N_{\Ns=2}q_m=m\,q_m\,,\notag\\
  &\eta=\sum_{m=0}^{\ol{p}}\eta_m\,,\quad N_{\Ns=2}\eta_m=m\,\eta_m
 \label{4D114} 
\end{align}
where $\eta_{\ol{p}}$ does not vanish and $p_{\ol{m}}$ or $q_{\ol{m}}$ do not vanish.
By arguments as in the text following equations \eqref{4D88} we can assume with no loss of generality that $\ol{m}-2\leq\ol{p}\leq\ol{m}$. 

In the case $\ol{p}=\ol{m}$, the coboundary condition \eqref{4D113a} yields at $N_{\Ns=2}$-eigenvalues $\ol{m}+2$ and $\ol{m}+1$
\begin{align}
  \fbb{\Ns=2}\eta_{\ol{m}}=0,\quad \fbb{\Ns=2}\eta_{\ol{m}-1}=0.
 \label{4D115} 
\end{align}
Any $\fbb{\Ns=2}$-exact contribution $\fbb{\Ns=2}\varrho_{\ol{p}-2}$ to $\eta_{\ol{p}}$ can be removed from $\eta$ by replacing $\eta$ with $\eta-\fb\varrho_{\ol{p}-2}$ as this replacement does not affect the coboundary condition \eqref{4D113a} (owing to $\fb^2=0$).
Therefore, using lemma \ref{lem4D6} and a notation as above, we infer from the first equation \eqref{4D115} that with no loss of generality we can assume
\begin{align}
  \eta_{\ol{m}}&=\7p_{\ol{m}}(\pb1,\cb1,\xi_2,\dots,\xi_\Ns)
  +\7q_{\ol{m}}(\psi_1,\chi_1,\xi_2,\dots,\xi_\Ns)\notag\\
  &+(-\tc2\pb1\psi_2+\tc4\cb1\psi_2-\tc1\cb1\chi_2+\tc3\pb1\chi_2)
  \7h_{\ol{m}-2}(\pb1,\cb1,\xi_2,\dots,\xi_\Ns)\notag\\
  &+(-\tc2\psi_1\pb2+\tc3\chi_1\pb2-\tc1\chi_1\cb2+\tc4\psi_1\cb2)
  \7g_{\ol{m}-2}(\psi_1,\chi_1,\xi_2,\dots,\xi_\Ns)\notag\\
  &+(\tc1\chi_1\cb1-\tc3\chi_1\pb1+\tc2\psi_1\pb1-\tc4\psi_1\cb1\notag\\
  &\phantom{+(}-\tc1\chi_2\cb2+\tc3\chi_2\pb2-\tc2\psi_2\pb2+\tc4\psi_2\cb2)\7b_{\ol{m}-2}(\xi_2,\dots,\xi_\Ns).
 \label{4D116} 
\end{align}
The second equation \eqref{4D115} implies an analogous result for $\eta_{\ol{m}-1}$.

At $N_{\Ns=2}$-eigenvalue $\ol{m}$, the coboundary condition \eqref{4D113a} yields in the case $\ol{p}=\ol{m}$
\begin{align}
  p_{\ol{m}}+q_{\ol{m}}=\fbb{\Ns=2}\eta_{\ol{m}-2}+\fbb{\Ns >2}\eta_{\ol{m}}\,.
 \label{4D117} 
\end{align}
Using the result \eqref{4D116} in equation \eqref{4D117}, one obtains analogously to equation \eqref{4D108}
\begin{align}
  p_{\ol{m}}+q_{\ol{m}}=&\,\fbb{\Ns=2}\eta_{\ol{m}-2}\notag\\
  &+\7h_{\ol{m}-2}(\pb1,\cb1,\xi_2,\dots,\xi_\Ns)
  \sum_{i=3}^\Ns(\pb{i}\cb1-\cb{i}\pb1)(\chi_i\psi_2-\psi_i\chi_2)
  \notag\\
  &+\7g_{\ol{m}-2}(\psi_1,\chi_1,\xi_2,\dots,\xi_\Ns)
  \sum_{i=3}^\Ns(\pb{i}\cb2-\cb{i}\pb2)(\chi_i\psi_1-\psi_i\chi_1)
  \notag\\
  &+\fbb{\Ns=2}\sum_{i=3}^\Ns\7b_{\ol{m}-2}(\xi_2,\dots,\xi_\Ns)
  (\tc1\chi_i\cb{i}-\tc3\chi_i\pb{i}+\tc2\psi_i\pb{i}-\tc4\psi_i\cb{i})\notag\\
  &-2\7b_{\ol{m}-2}(\xi_2,\dots,\xi_\Ns)
  \sum_{i=3}^\Ns(\pb{i}\cb2-\cb{i}\pb2)(\psi_i\chi_2-\chi_i\psi_2).
 \label{4D118} 
\end{align}
We write this equation as
\begin{align}
  p^\prime_{\ol{m}}(\pb1,\cb1,\xi_2,\dots,\xi_\Ns)
  +q^\prime_{\ol{m}}(\psi_1,\chi_1,\xi_2,\dots,\xi_\Ns)
  +b^\prime_{\ol{m}}(\xi_2,\dots,\xi_\Ns)
  =\fbb{\Ns=2}\eta^\prime_{\ol{m}-2}
 \label{4D120} 
\end{align}
where, leaving out the arguments of $p^\prime_{\ol{m}}$, $q^\prime_{\ol{m}}$ and $b^\prime_{\ol{m}}$,
\begin{align}
  p^\prime_{\ol{m}}&=
  p_{\ol{m}}-\7h_{\ol{m}-2}(\pb1,\cb1,\xi_2,\dots,\xi_\Ns)
  \sum_{i=3}^\Ns(\pb{i}\cb1-\cb{i}\pb1)(\chi_i\psi_2-\psi_i\chi_2)
  \notag\\
  q^\prime_{\ol{m}}&=
  q_{\ol{m}}-\7g_{\ol{m}-2}(\psi_1,\chi_1,\xi_2,\dots,\xi_\Ns)
  \sum_{i=3}^\Ns(\pb{i}\cb2-\cb{i}\pb2)(\chi_i\psi_1-\psi_i\chi_1)
  \notag\\
  b^\prime_{\ol{m}}&=2\7b_{\ol{m}-2}(\xi_2,\dots,\xi_\Ns)
  \sum_{i=3}^\Ns(\pb{i}\cb2-\cb{i}\pb2)(\psi_i\chi_2-\chi_i\psi_2)
  \notag\\
  \eta^\prime_{\ol{m}-2}&=
  \eta_{\ol{m}-2}
  +\sum_{i=3}^\Ns\7b_{\ol{m}-2}(\xi_2,\dots,\xi_\Ns)
  (\tc1\chi_i\cb{i}-\tc3\chi_i\pb{i}+\tc2\psi_i\pb{i}-\tc4\psi_i\cb{i}).
 \label{4D121} 
\end{align}
According to equation \eqref{4D120}, $p^\prime_{\ol{m}}+q^\prime_{\ol{m}}+b^\prime_{\ol{m}}$ is a sum of polynomials in the supersymmetry ghosts which either do not depend on $\psi_1$ and $\chi_1$ or on $\pb1$ and $\cb1$. Using lemma \ref{lem4D7}, we conclude from equation \eqref{4D120}
\begin{align}
   &p^\prime_{\ol{m}}(\pb1,\cb1,\xi_2,\dots,\xi_\Ns)
  +q^\prime_{\ol{m}}(\psi_1,\chi_1,\xi_2,\dots,\xi_\Ns)
  +b^\prime_{\ol{m}}(\xi_2,\dots,\xi_\Ns)\notag\\
   =&\,(\pb2\cb1-\cb2\pb1)(\chi_2p^\prime_{2,\ol{m}-3}(\pb1,\cb1,\xi_2,\dots,\xi_\Ns)
   +\psi_2p^\prime_{3,\ol{m}-3}(\pb1,\cb1,\xi_2,\dots,\xi_\Ns))\notag\\
   &+(\psi_2\chi_1-\chi_2\psi_1)(\cb2q^\prime_{2,\ol{m}-3}(\psi_1,\chi_1,\xi_2,\dots,\xi_\Ns)
   -\pb2q^\prime_{3,\ol{m}-3}(\psi_1,\chi_1,\xi_2,\dots,\xi_\Ns)).
  \label{4D122} 
\end{align}
Equations \eqref{4D121} and \eqref{4D122} yield
\begin{align}
  &p_{\ol{m}}+q_{\ol{m}}=
  -2\7b_{\ol{m}-2}(\xi_2,\dots,\xi_\Ns)
  \sum_{i=3}^\Ns(\pb{i}\cb2-\cb{i}\pb2)(\psi_i\chi_2-\chi_i\psi_2)
  \notag\\
  &+(\pb2\cb1-\cb2\pb1)(\chi_2p^\prime_{2,\ol{m}-3}(\pb1,\cb1,\xi_2,\dots,\xi_\Ns)
   +\psi_2p^\prime_{3,\ol{m}-3}(\pb1,\cb1,\xi_2,\dots,\xi_\Ns))
  \notag\\
  &+(\psi_2\chi_1-\chi_2\psi_1)(\cb2q^\prime_{2,\ol{m}-3}(\psi_1,\chi_1,\xi_2,\dots,\xi_\Ns)
   -\pb2q^\prime_{3,\ol{m}-3}(\psi_1,\chi_1,\xi_2,\dots,\xi_\Ns))
   \notag\\
  &+\7h_{\ol{m}-2}(\pb1,\cb1,\xi_2,\dots,\xi_\Ns)
  \sum_{i=3}^\Ns(\pb{i}\cb1-\cb{i}\pb1)(\chi_i\psi_2-\psi_i\chi_2)
  \notag\\
  &+\7g_{\ol{m}-2}(\psi_1,\chi_1,\xi_2,\dots,\xi_\Ns)
  \sum_{i=3}^\Ns(\pb{i}\cb2-\cb{i}\pb2)(\chi_i\psi_1-\psi_i\chi_1)
  \notag\\
  &=\,\fb\4\eta
  +\7b_{\ol{m}-2}(\xi_2,\dots,\xi_\Ns)
  \sum_{i=3}^\Ns\sum_{j=3}^\Ns(\pb{i}\cb{j}-\cb{i}\pb{j})(\psi_i\chi_j-\chi_i\psi_j)
  \notag\\
  &-\sum_{i=3}^\Ns\Big[(\psi_i\chi_1-\chi_i\psi_1)(\cb{i}q^\prime_{2,\ol{m}-3}(\psi_1,\chi_1,\xi_2,\dots,\xi_\Ns)
   -\pb{i}q^\prime_{3,\ol{m}-3}(\psi_1,\chi_1,\xi_2,\dots,\xi_\Ns))
  \notag\\
  &+(\pb{i}\cb1-\cb{i}\pb1)(\chi_ip^\prime_{2,\ol{m}-3}(\pb1,\cb1,\xi_2,\dots,\xi_\Ns)
   +\psi_ip^\prime_{3,\ol{m}-3}(\pb1,\cb1,\xi_2,\dots,\xi_\Ns))\Big]
  \label{4D223a}
\end{align}
where, leaving out the arguments of $\7b_{\ol{m}-2}$ etc.,
\begin{align}
  \4\eta=&\,
  \Big[(\tc1\cb{1}\chi_1-\tc3\pb{1}\chi_1+\tc2\pb{1}\psi_1-\tc4\cb{1}\psi_1)
  \notag\\
  &-\sum_{i=2}^\Ns(\tc1\cb{i}\chi_i-\tc3\pb{i}\chi_i+\tc2\pb{i}\psi_i-\tc4\cb{i}\psi_i)\Big]
  \7b_{\ol{m}-2}
  \notag\\
  &+(\tc4\cb1-\tc2\pb1)(p^\prime_{2,\ol{m}-3}+\psi_2\7h_{\ol{m}-2})
  +(\tc1\cb1-\tc3\pb1)(p^\prime_{3,\ol{m}-3}-\chi_2\7h_{\ol{m}-2})\notag\\
  &+(\tc3\chi_1-\tc2\psi_1)(q^\prime_{2,\ol{m}-3}+\pb2\7g_{\ol{m}-2})
  +(\tc4\psi_1-\tc1\chi_1)(q^\prime_{3,\ol{m}-3}+\cb2\7g_{\ol{m}-2}).
  \label{4D223b}
\end{align}
$\4\eta$ contributes to $\7\eta$ in \eqref{4D224} at $\Ns_{\Ns=2}$-degrees $\ol{m}$ and $\ol{m}-2$, with $p^\prime_{2,\ol{m}-3}+\psi_2\7h_{\ol{m}-2}$ contributing to $\4p_2$ et cetera.
If $\ol{p}=\ol{m}-2$ or $\ol{p}=\ol{m}-1$, the coboundary condition \eqref{4D113a} yields
$p_{\ol{m}}+q_{\ol{m}}=\fbb{\Ns=2}\eta_{\ol{m}-2}$ and, in the case $\ol{p}=\ol{m}-1$, additionally $\fbb{\Ns=2}\eta_{\ol{m}-1}=0$. The latter are equations as \eqref{4D115} and \eqref{4D117} with $\eta_{\ol{m}}=0$ and lead to results for $p_{\ol{m}}+q_{\ol{m}}$ as in equations \eqref{4D223a} and \eqref{4D223b} with $\7h_{\ol{m}-2}=\7g_{\ol{m}-2}=\7b_{\ol{m}-2}=0$. 

Equation \eqref{4D223a} shows that $p(\pb1,\cb1,\xi_2,\dots,\xi_\Ns)+q(\psi_1,\chi_1,\xi_2,\dots,\xi_\Ns)-\fb\eta^\prime$ is an $\fb$-exact polynomial in the supersymmetry ghosts of the form $p'(\pb1,\cb1,\xi_2,\dots,\xi_\Ns)+q'(\psi_1,\chi_1,\xi_2,\dots,\xi_\Ns)$ which contains no terms with $N_{\Ns=2}$-eigenvalues $m\geq\ol{m}$. Hence, it can be treated as $p(\pb1,\cb1,\xi_2,\dots,\xi_\Ns)+q(\psi_1,\chi_1,\xi_2,\dots,\xi_\Ns)$ above and the process can be continued until the $N_{\Ns=2}$-eigenvalue drops to zero which yields lemma \ref{lem4D10}. \QED

\subsection{\texorpdfstring{$\Hg(\fb)$ in covariant form}{H(gh) in covariant form}}\label{4D.2}

We shall now provide \so{t,4-t}-covariant versions of the results for $\Dim=4$ which extend these results to all spinor rerpesentations equivalent to the particular representations \eqref{4D1}. To this end we introduce the following \so{t,4-t}-covariant ghost polynomials (with $\xi_i^\pm=\half\xi_i(\unit\pm\GAM)$):
\begin{align}
  \vartheta_i^\ua
  &=c^a\xi^\ub_i\gam_{a\ub}{}^\ua,\notag\\
  \vartheta_i^{+\ua}&=\half\vartheta_i^\ub(\unit+\GAM)_\ub{}^\ua=c^a\xi^{-\ub}_i\gam_{a\ub}{}^\ua,\notag\\
  \vartheta_i^{-\ua}&=\half\vartheta_i^\ub(\unit-\GAM)_\ub{}^\ua=c^a\xi^{+\ub}_i\gam_{a\ub}{}^\ua,\notag\\
  \Theta_{ij}&=\vartheta_i^{+}\cdot\xi_j^+
  =\quart\vartheta_i^\ug(\unit+\GAM)_\ug{}^\ua\xi_j^\ud(\unit+\GAM)_\ud{}^\ub\IC_{\ua\ub}
  =c^a\xi_i^\ua(\half\gam_a(\unit+\GAM)\IC)_{\ua\ub}\xi_j^\ub
  \label{4D7}
\end{align}
where $\vartheta_i^{+}\cdot\xi_j^+$ denotes the \so{t,4-t}-invariant product $\vartheta_i^{+\ua}\IC_{\ua\ub}\xi_j^{+\ub}$ of $\vartheta_i^{+}$ and $\xi_j^+$ (cf. section 2.6 of \pap).
We note that the products $\vartheta_i^{-}\cdot\xi_j^-$ can be expressed in terms of the products $\vartheta_i^{+}\cdot\xi_j^+$ (and vice versa):
\begin{align}
\vartheta_i^{+}\cdot\xi_j^+
  &=c^a\xi_i^\ua(\half\gam_a(\unit+\GAM)\IC)_{\ua\ub}\xi_j^\ub\notag\\
  &=c^a\xi_i^\ua(\half\gam_a(\unit-\GAM)\IC)_{\ub\ua}\xi_j^\ub
  =\vartheta_j^{-}\cdot\xi_i^-.
 \label{4D8}
\end{align}

The coboundary operator $\fb$ acts on the $\vartheta_i^\pm$ and $\Theta_{ij}$ according to
\begin{align}
&\fb\vartheta_i^{+\ua}=2\Ii\sum_{j=1}^\Ns(\xi^-_i\cdot\xi^-_j)\, \xi^{+\ua}_j\, ,\notag\\
&\fb\vartheta_i^{-\ua}=2\Ii\sum_{j=1}^\Ns(\xi^+_i\cdot\xi^+_j)\, \xi^{-\ua}_j\, ,\notag\\
&\fb\Theta_{ij}=-2\Ii\sum_{k=1}^\Ns(\xi^-_i\cdot\xi^-_k)(\xi^+_j\cdot\xi^+_k).
\label{4D9}
\end{align}

In the spinor representations \eqref{4D1} one has:
\begin{align}
  &(\vartheta_i^{\ul1},\vartheta_i^{\ul2},\vartheta_i^{\ul3},\vartheta_i^{\ul4})=
  2(\tc1\cb{i}-\tc3\pb{i},-\tc1\chi_i+\tc4\psi_i,-\tc2\psi_i+\tc3\chi_i,-\tc2\pb{i}+\tc4\cb{i}),
  \notag\\
  &(\vartheta_i^{+\ul1},\vartheta_i^{+\ul2},\vartheta_i^{+\ul3},\vartheta_i^{+\ul4})
  =(\vartheta_i^{\ul1},0,0,\vartheta_i^{\ul4}),\quad
  (\vartheta_i^{-\ul1},\vartheta_i^{-\ul2},\vartheta_i^{-\ul3},\vartheta_i^{-\ul4})
  =(0,\vartheta_i^{\ul2},\vartheta_i^{\ul3},0),
  \notag\\
  &\Theta_{ij}=2\Ii(-\tc1\cb{i}\chi_j+\tc3\pb{i}\chi_j-\tc2\pb{i}\psi_j+\tc4\cb{i}\psi_j),
  \notag\\
  &\xi^+_i\cdot\xi^+_j=-\Ii\psi_i\chi_j+\Ii\chi_i\psi_j,\quad
  \xi^-_i\cdot\xi^-_j=\Ii\pb{i}\cb{j}-\Ii\cb{i}\pb{j},\notag\\
  &\vartheta_1^+\cdot \vartheta_1^+=8\Ii(\tc1\cb1-\tc3\pb1)(\tc2\pb1-\tc4\cb1),\notag\\
  &\vartheta_1^-\cdot \vartheta_1^-=8\Ii(\tc1\chi_1-\tc4\psi_1)(\tc3\chi_1-\tc2\psi_1).
  \label{4D7a}
\end{align}
Using these expressions one straightforwardly verifies equations \eqref{4D9} in the spinor representations \eqref{4D1} which implies that they also hold in any spinor representation equivalent to \eqref{4D1} owing to their \so{t,4-t}-covariance. Furthermore these expressions show that various ghost polynomials in lemmas \ref{lem4D4} to \ref{lem4D10} can be expressed in an \so{t,4-t}-covariant way. Using additionally that equivalence transformations relating equivalent spinor representations do not mix chiralities of spinors in the sense of section 2.7 of \pap, one can directly  obtain from lemmas \ref{lem4D4} to \ref{lem4D10} the following results that are valid for all spinor representations equivalent to \eqref{4D1}.

The covariant version of lemma \ref{lem4D4} is:

\begin{lemma}[$\Hg(\fb)$ for $\Ns=1$]\label{lem4D5}\quad \\
In the case $\Ns=1$\\
(i) any cocycle in $\Hg(\fb)$ is equivalent to a linear combination of a polynomial in the components of $\xi^-_1$ and
$\vartheta^+_1$, of a polynomial in the components of $\xi^+_1$ and
$\vartheta^-_1$, and of $\Theta_{11}$, with $\vartheta^\pm_1$ and $\Theta_{11}$ as in equations \eqref{4D7}:
\begin{align}
  \fb\om=0\ \LRA\ 
  \om\sim&\,
  \Theta_{11}b+p(\xi^-_1)+q(\xi^+_1)
  +\vartheta^{+\ua}_1 p_\ua(\xi^-_1)
  +\vartheta^{-\ua}_1 q_\ua(\xi^+_1)\notag\\
  &+(\vartheta_1^+\cdot \vartheta_1^+)p_-(\xi^-_1)
  +(\vartheta_1^-\cdot \vartheta_1^-)q_+(\xi^+_1)
  \label{4D67}
\end{align}
with arbitrary polynomials $p(\xi^-_1)$, $p_\ua(\xi^-_1)$, $p_-(\xi^-_1)$ in the components of $\xi^-_1$,
arbitrary polynomials $q(\xi^+_1)$, $q_\ua(\xi^+_1)$, $q_+(\xi^+_1)$ in the components of $\xi^+_1$,
and an arbitrary complex number $b\in\mathbb{C}$;\\
(ii) a linear combination of a polynomial in the components of $\xi^-_1$ and
$\vartheta^+_1$, of a polynomial in the components of $\xi^+_1$ and
$\vartheta^-_1$, and of $\Theta_{11}$ is exact in $\Hg(\fb)$ if and only if it vanishes:
\begin{align}
  0\sim
  &\,\Theta_{11}b+p(\xi^-_1)+q(\xi^+_1)
  +\vartheta^{+\ua}_1 p_\ua(\xi^-_1)
  +\vartheta^{-\ua}_1 q_\ua(\xi^+_1)\notag\\
  &+(\vartheta_1^+\cdot \vartheta_1^+)p_-(\xi^-_1)
  +(\vartheta_1^-\cdot \vartheta_1^-)q_+(\xi^+_1)
  \notag\\
  \LRA\quad&
  b=p(\xi^-_1)+q(\xi^+_1)=\vartheta^{+\ua}_1 p_\ua(\xi^-_1)=
  \vartheta^{-\ua}_1 q_\ua(\xi^+_1)=p_-(\xi^-_1)=q_+(\xi^+_1)=0.
  \label{4D68}
\end{align}
\end{lemma}

{\bf Comments}:

1. Lemma \ref{lem4D5} reproduces for signatures $(1,3)$ and $(3,1)$ the results derived in section 13.1 of \cite{Brandt:1991as} and in \cite{Dixon:1993jt} when particularized for the spinor representations considered there.

2. In the case $\Ns=1$ equations \eqref{4D9} yield $\fb\vartheta_1^{+\,\ua}=2\Ii (\xi_1^-\cdot\xi_1^-)\xi_1^{+\,\ua}=0$ (owing to $\xi_i^-\cdot\xi_j^-=-\xi_j^-\cdot\xi_i^-$ which implies $\xi_1^-\cdot\xi_1^-=0$) and analogously $\fb\vartheta_1^{-\,\ua}=0$ as well as $\fb\Theta_{11}=-2\Ii(\xi_1^-\cdot\xi_1^-)(\xi_1^+\cdot\xi_1^+)=0$ which shows that $\vartheta^{+\ua}_1$, $\vartheta^{-\ua}_1$ and $\Theta_{11}$ are indeed cocycles in $\Hg(\fb)$ in the case $\Ns=1$.

3. The cocycles $p(\xi^-_1)$, $\vartheta^{+\ua}_1 p_\ua(\xi^-_1)$ and $(\vartheta_1^+\cdot \vartheta_1^+)p_-(\xi^-_1)$ depend on the supersymmetry ghosts only via components $\xi_1^{-\, \ua}$ of negative chirality since $\vartheta^{+}_1$ also depends on the supersymmetry ghosts only via the $\xi_1^{-\, \ua}$. Analogously the cocycles $q(\xi^+_1)$, $\vartheta^{-\ua}_1 q_\ua(\xi^+_1)$ and $(\vartheta_1^-\cdot \vartheta_1^-)q_+(\xi^+_1)$ depend on the supersymmetry ghosts only via components $\xi_1^{+\,\ua}$ of positive chirality.  In contrast, $\Theta_{11}$ depends linearly both on components $\xi_1^{-\, \ua}$ with negative chirality and on components $\xi_1^{+\,\ua}$ with positive chirality.

Lemmas \ref{lem4D6} and \ref{lem4D7} yield:

\begin{lemma}[$\Hg(\fb)$ for $\Ns=2$]\label{lem4D8}\quad\\
In the case $\Ns=2$\\
(i) the general solution of the cocycle condition in $\Hg(\fb)$ is
\begin{align}
 \fb\om=0\ \LRA\ \om\sim&\,P(\xi_1^-,\xi_2)+Q(\xi_1^+,\xi_2)
  +\Theta_{12}H(\xi_1^-,\xi_2)\notag\\
  &+\Theta_{21}G(\xi_1^+,\xi_2)
  +(\Theta_{11}-\Theta_{22})B(\xi_2)
 \label{4D97} 
\end{align}
with $\Theta_{ij}$ as in equations \eqref{4D7}, arbitrary polynomials
$P(\xi_1^-,\xi_2)$, $H(\xi_1^-,\xi_2)$ in the components of $\xi_1^-$ and $\xi_2$, arbitrary polynomials $Q(\xi_1^+,\xi_2)$, $G(\xi_1^+,\xi_2)$ in the components of $\xi_1^+$ and $\xi_2$, and an arbitrary polynomial $B(\xi_2)$ in the components of $\xi_2$;

(ii) a cocycle $P(\xi_1^-,\xi_2)+Q(\xi_1^+,\xi_2)+\Theta_{12}H(\xi_1^-,\xi_2)+\Theta_{21}G(\xi_1^+,\xi_2)
+(\Theta_{11}-\Theta_{22})B(\xi_2)$ is $\fb$-exact if and only if $P(\xi_1^-,\xi_2)+Q(\xi_1^+,\xi_2)$ is a linear combination of polynomials $(\xi_1^-\cdot\xi_2^-)\xi_2^{+\ua}\4P_\ua(\xi_1^-,\xi_2)$ and $(\xi_1^+\cdot\xi_2^+)\xi_2^{-\ua}\4Q_\ua(\xi_1^+,\xi_2)$ and $H(\xi_1^-,\xi_2)$ and $G(\xi_1^+,\xi_2)$ factorize according to $(\xi_1^-\cdot\xi_2^-)\4P(\xi_1^-,\xi_2)$ and $(\xi_1^+\cdot\xi_2^+)\4Q(\xi_1^+,\xi_2)$, respectively, and $B(\xi_2)$ vanishes:
\begin{align}
  0\sim &\ P(\xi_1^-,\xi_2)+Q(\xi_1^+,\xi_2)
  +\Theta_{12}H(\xi_1^-,\xi_2)+\Theta_{21}G(\xi_1^+,\xi_2)
  +(\Theta_{11}-\Theta_{22})B(\xi_2)\notag\\
  \LRA\ &P(\xi_1^-,\xi_2)+Q(\xi_1^+,\xi_2)=(\xi_1^-\cdot\xi_2^-)\xi_2^{+\ua}\4P_\ua(\xi_1^-,\xi_2)
  +(\xi_1^+\cdot\xi_2^+)\xi_2^{-\ua}\4Q_\ua(\xi_1^+,\xi_2)\ \wedge\notag\\
 & H(\xi_1^-,\xi_2)=(\xi_1^-\cdot\xi_2^-)\4P(\xi_1^-,\xi_2)\ \wedge\notag\\
 &G(\xi_1^+,\xi_2)=(\xi_1^+\cdot\xi_2^+)\4Q(\xi_1^+,\xi_2)\ \wedge\notag\\
 &B(\xi_2)=0.
 \label{4D98} 
\end{align}
\end{lemma}

{\bf Comments:}

4. The third equation \eqref{4D9} yields 
\begin{align}
\Ns=2:\quad &\fb\Theta_{12}=0=\fb\Theta_{21}\,,\notag\\
&\fb\Theta_{11}=-2\Ii(\xi_1^-\cdot\xi_2^-)\,(\xi_1^+\cdot\xi_2^+)=\fb\Theta_{22}\,.\label{4D99}
\end{align}
This verifies that the terms in \eqref{4D97} are indeed $\fb$-closed in the case $\Ns=2$.

5. The first and second equation \eqref{4D9} and equations \eqref{4D7} and \eqref{4D8} yield 
\begin{align}
\Ns=2:\quad &(\xi_1^-\cdot\xi_2^-)\xi_2^{+\ua}=-\ihalf\,\fb\vartheta_1^{+\,\ua},\notag\\
&(\xi_1^+\cdot\xi_2^+)\xi_2^{-\ua}=-\ihalf\,\fb\vartheta_1^{-\,\ua},\notag\\
&(\xi_1^-\cdot\xi_2^-)\Theta_{12}=(\xi_1^-\cdot\xi_2^-)\,(\vartheta_1^+\cdot\xi_2^+)
=-\ihalf\,(\vartheta_1^+\cdot\fb\vartheta_1^+)
=\ifourth\,\fb(\vartheta_1^+\cdot\vartheta_1^+),\notag\\
&(\xi_1^+\cdot\xi_2^+)\Theta_{21}=(\xi_1^+\cdot\xi_2^+)\,(\vartheta_1^-\cdot\xi_2^-)
=-\ihalf\,(\vartheta_1^-\cdot\fb\vartheta_1^-)
=\ifourth\,\fb(\vartheta_1^-\cdot\vartheta_1^-).\label{4D100}
\end{align}
These relations underlie the result \eqref{4D98}.

Lemmas \ref{lem4D9} and \ref{lem4D10} yield:

\begin{lemma}[$\Hg(\fb)$ for $\Ns>2$]\label{lem4D11}\quad\\
In the cases $\Ns>2$\\
(i) the general solution of the cocycle condition in $\Hg(\fb)$ is
\begin{align}
  \fb\om=0\ \LRA\ \om\sim P(\xi_1^-,\xi_2,\dots,\xi_\Ns)+Q(\xi_1^+,\xi_2,\dots,\xi_\Ns)
 \label{4D225} 
\end{align}
with an arbitrary polynomial $P(\xi_1^-,\xi_2,\dots,\xi_\Ns)$ in the components of $\xi_1^-$, $\xi_2$, \dots , $\xi_\Ns$ and an arbitrary polynomial $Q(\xi_1^+,\xi_2,\dots,\xi_\Ns)$ in the components of $\xi_1^+$, $\xi_2$, \dots , $\xi_\Ns$;

(ii) a cocycle $P(\xi_1^-,\xi_2,\dots,\xi_\Ns)+Q(\xi_1^+,\xi_2,\dots,\xi_\Ns)$ is $\fb$-exact if and only if it is the $\fb$-transformation of a linear combination of $\Theta_{11}-\sum_{i=2}^\Ns\Theta_{ii}$ times a polynomial in the components of $\xi_2$, \dots , $\xi_\Ns$, of the components of $\vartheta_1^{+}$ times polynomials in the components of $\xi_1^-$, $\xi_2$, \dots , $\xi_\Ns$, and of the components of $\vartheta_1^{-}$ times polynomials in the components of $\xi_1^+$, $\xi_2$, \dots , $\xi_\Ns$:
\begin{align}
  &P(\xi_1^-,\xi_2,\dots,\xi_\Ns)+Q(\xi_1^+,\xi_2,\dots,\xi_\Ns)\sim 0\notag\\
  \LRA\ &P(\xi_1^-,\xi_2,\dots,\xi_\Ns)+Q(\xi_1^+,\xi_2,\dots,\xi_\Ns)\notag\\
  &=\fb\Big[
  (\Theta_{11}-\sum_{i=2}^\Ns\Theta_{ii})\7B(\xi_2,\dots,\xi_\Ns)
  +\vartheta_1^{+\,\ua}\4P_\ua(\xi_1^-,\xi_2,\dots,\xi_\Ns)\notag\\
  &\phantom{=\fb\Big[}
  +\vartheta_1^{-\,\ua}\4Q_\ua(\xi_1^+,\xi_2,\dots,\xi_\Ns)
  \Big]
  \label{4D226}\\
  &=2\Ii\sum_{i=2}^\Ns\sum_{j=2}^\Ns(\xi_i^+\cdot\xi_j^+)(\xi_i^-\cdot\xi_j^-)\7B(\xi_2,\dots,\xi_\Ns)
  \notag\\
  &\phantom{=}+2\Ii\sum_{i=2}^\Ns\Big[
  (\xi_1^-\cdot\xi_i^-)\xi_i^{+\ua}\4P_\ua(\xi_1^-,\xi_2,\dots,\xi_\Ns)
  +(\xi_1^+\cdot\xi_i^+)\xi_i^{-\ua}\4Q_\ua(\xi_1^+,\xi_2,\dots,\xi_\Ns)\Big].
 \label{4D227}
\end{align}
\end{lemma}

{\bf Comment:}

6. One has
\[
\fb(\Theta_{11}-\sum_{i=2}^\Ns\Theta_{ii})=
2\Ii\sum_{i=2}^\Ns\sum_{j=2}^\Ns(\xi_i^+\cdot\xi_j^+)(\xi_i^-\cdot\xi_j^-).
\]
This shows that for $\Ns>2$ there are polynomials in the supersymmetry ghosts which do not depend on components of $\xi_1$ and are nevertheless $\fb$-exact. According to part (ii) of lemma \ref{lem4D11}, these polynomials are of the form $\fb(\Theta_{11}-\sum_{i=2}^\Ns\Theta_{ii})\7B(\xi_2,\dots,\xi_\Ns)$.

\section{Primitive elements in five dimensions}\label{5D}

\subsection{\texorpdfstring{Relation of $\fb$-transformations in $\Dim=4$ and $\Dim=5$}{Relation of transformations in D=4 and D=5}}\label{5D.0}

We shall use the results in $\Dim=4$ dimensions to derive the results in $\Dim=5$ dimensions. To this end we first relate the $\fb$-transformations in $\Dim=4$ and $\Dim=5$ by marking $\Dim=4$ objects by a subscript $(\Dim=4)$. For spinor representations in $\Dim=4$ and $\Dim=5$ with the same gamma-matrices $\gam^1,\dots,\gam^4$, the $\Dim=5$ charge conjugation matrix $\CC$ and $\gam_5$ are related to the $\Dim=4$ charge conjugation matrix $\CC_{(\Dim=4)}$ and $\GAM_{(\Dim=4)}$, respectively, by: 
\begin{align}
  \CC=\CC_{(\Dim=4)}\GAM_{(\Dim=4)}\, ,\quad \gam_5=k_5 \GAM_{(\Dim=4)}.
  \label{5D0}
\end{align}
Decomposing the $\Dim=5$ supersymmetry ghosts according to
\begin{align}
  \xi_i=\xi_i^++\xi_i^-,\quad \xi_i^\pm=\half\, \xi_i(\unit\pm\GAM_{(\Dim=4)}),
  \label{5D1}
\end{align}
we have in $\Dim=5$, using matrix notation with $\xi_i=(\xx1_i,\xx2_i,\xx3_i,\xx4_i)$:
\begin{align}
  a\in\{1,\dots,4\}:\ \fb c^a&=\Ii\sum_{k=1}^{\Ns/2}\xi_{2k-1}\gam^a\IC(\xi_{2k})^\top\notag\\
                                 &=\Ii\sum_{k=1}^{\Ns/2}\Big(\xi_{2k-1}^-\gam^a\IC_{(\Dim=4)}(\xi_{2k}^+)^\top-
                                                         \xi_{2k-1}^+\gam^a\IC_{(\Dim=4)}(\xi_{2k}^-)^\top\Big).
  \label{5D2}
\end{align}
In $\Dim=4$ we have:
\begin{align}
  \fb c^a_{(\Dim=4)}=\ihalf\sum_{i=1}^{\Ns}\xi_{i(\Dim=4)}\gam^a\IC_{(\Dim=4)}(\xi_{i(\Dim=4)})^\top
  =\Ii\sum_{i=1}^{\Ns}\xi_{i(\Dim=4)}^+\gam^a\IC_{(\Dim=4)}(\xi_{i(\Dim=4)}^-)^\top.
  \label{5D3}
\end{align}
Comparing \eqref{5D2} and \eqref{5D3} we observe that we can match the $\fb$-transformations of $c^1,\dots,c^4$ in $\Dim=4$ and $\Dim=5$ for $\Ns\in\{2,4,6,\dots\}$ using the identifications
\begin{align}
   a=1,\dots,4:\ &c^a= c^a_{(\Dim=4)}\, ;\notag\\
  k=1,\dots,\frac{\Ns}2:\ &\xi_{2k-1}^+= \xi_{2k-1(\Dim=4)}^+,\
  \xi_{2k-1}^-= \xi_{2k(\Dim=4)}^-,\ \xi_{2k-1}= \xi_{2k-1(\Dim=4)}^++\xi_{2k(\Dim=4)}^-,\notag\\
  &\xi_{2k}^+= \xi_{2k(\Dim=4)}^+,\
  \xi_{2k}^-= -\xi_{2k-1(\Dim=4)}^-,\ \xi_{2k}= \xi_{2k(\Dim=4)}^+-\xi_{2k-1(\Dim=4)}^-.
  \label{5D4}
\end{align}
With these identifications we obtain
\begin{align}
   \fb c^5=&\Ii\sum_{k=1}^{\Ns/2}\xi_{2k-1}\gam^5\IC(\xi_{2k})^\top
   =\frac{\Ii}{k_5}\sum_{k=1}^{\Ns/2}\xi_{2k-1}\IC_{(\Dim=4)}(\xi_{2k})^\top\notag\\
   =&\frac{\Ii}{k_5}\sum_{k=1}^{\Ns/2}\Big(\xi_{2k-1}^+\IC_{(\Dim=4)}(\xi_{2k}^+)^\top
                              +\xi_{2k-1}^-\IC_{(\Dim=4)}(\xi_{2k}^-)^\top\Big)\notag\\
   =&\frac{\Ii}{k_5}\sum_{k=1}^{\Ns/2}\Big[\xi_{2k-1}^+\IC(\xi_{2k}^+)^\top
                              -\xi_{2k}^-\IC(\xi_{2k-1}^-)^\top\Big]_{(\Dim=4)}\notag\\
   =&\frac{\Ii}{k_5}\sum_{k=1}^{\Ns/2}[\xi_{2k-1}^+\cdot\xi_{2k}^+
                              +\xi_{2k-1}^-\cdot\xi_{2k}^-]_{(\Dim=4)}
   =\frac{\Ii}{k_5}\sum_{k=1}^{\Ns/2}[\xi_{2k-1}\cdot\xi_{2k}]_{(\Dim=4)}
  \label{5D4a}
\end{align}
with all terms within parantheses $[\dots]_{(\Dim=4)}$ referring to $\Dim=4$. In particular, spinor products within such parantheses, such as $[\xi_{2k-1}\cdot\xi_{2k}]_{(\Dim=4)}$, refer to $\Dim=4$ spinor products of $\Dim=4$ spinors, i.e.
\[
[\xi_{2k-1}\cdot\xi_{2k}]_{(\Dim=4)}=\xi_{2k-1(\Dim=4)}\IC_{(\Dim=4)}(\xi_{2k(\Dim=4)})^\top.
\]


\subsection{\texorpdfstring{$\Hg(\fb)$ for $\Ns=2$}{H(gh) for N=2}}\label{5D.1}

We define \so{t,5-t}-covariant ghost polynomials $\theta_{ij}$ and $\theta_{a\,ij}$ according to
\begin{align}
  \theta_{ij}=c^a c^b\, \xi_i\,\gam_{ab}\IC\xi_j^\top ,\quad
  \theta_{a\,ij}=c^b \, \xi_i\,\gam_{ba}\IC\xi_j^\top 
	\label{5D8}
\end{align}
with
\begin{align}
  \gam_{ab}=\half(\gam_a\gam_b-\gam_b\gam_a).
	\label{5D9} 
\end{align}
In terms of $\Dim=4$ objects, one has for $i,j\in\{1,2\}$:
\begin{align}
  &\theta_{11}=-4k_5\,c^5\,\Theta_{21(\Dim=4)}
  +[\vartheta_2^+\cdot\vartheta_2^+-\vartheta_1^-\cdot\vartheta_1^-]_{(\Dim=4)}\,,\label{5D10}\\
  &\theta_{22}=4k_5\,c^5\,\Theta_{12(\Dim=4)}
  +[\vartheta_1^+\cdot\vartheta_1^+-\vartheta_2^-\cdot\vartheta_2^-]_{(\Dim=4)}\,,\label{5D11}\\
  &\theta_{12}=2k_5\,c^5\,[\Theta_{11}-\Theta_{22}]_{(\Dim=4)}-
  [\vartheta_1^-\cdot\vartheta_2^-+\vartheta_1^+\cdot\vartheta_2^+]_{(\Dim=4)}\,,\label{5D12}\\
  &\theta_{5\,11}=2k_5\,\Theta_{21(\Dim=4)}\,,\label{5D13}\\
  &\theta_{5\,22}=-2k_5\,\Theta_{12(\Dim=4)}\,,\label{5D14}\\
  &\theta_{5\,12}=k_5\,[\Theta_{22}-\Theta_{11}]_{(\Dim=4)}\,,\label{5D15}\\
&a\neq 5:\notag\\
  &\theta_{a\,11}=-2k_5\,c^5\,[\xi_1^+\gam_a\IC(\xi_2^-)^\top]_{(\Dim=4)}
  +[\vartheta_1^-\gam_a\IC(\xi_1^+)^\top-\vartheta_2^+\gam_a\IC(\xi_2^-)^\top]_{(\Dim=4)}
  \,,\label{5D16}\\
  &\theta_{a\,22}=2k_5\,c^5\,[\xi_2^+\gam_a\IC(\xi_1^-)^\top]_{(\Dim=4)}
  +[\vartheta_2^-\gam_a\IC(\xi_2^+)^\top-\vartheta_1^+\gam_a\IC(\xi_1^-)^\top]_{(\Dim=4)}
  \, ,\label{5D17}\\
  &\theta_{a\,12}=k_5\,c^5\,[\xi_1^+\gam_a\IC(\xi_1^-)^\top-\xi_2^+\gam_a\IC(\xi_2^-)^\top]_{(\Dim=4)}
  +\half[\vartheta_1^-\gam_a\IC(\xi_2^+)^\top\notag\\
  &\phantom{\theta_{a\,12}=}+\vartheta_2^+\gam_a\IC(\xi_1^-)^\top
  +\vartheta_2^-\gam_a\IC(\xi_1^+)^\top+\vartheta_1^+\gam_a\IC(\xi_2^-)^\top]_{(\Dim=4)}
  \label{5D18}
\end{align}
with $\Theta_{ij}$ and $\vartheta_i^\pm$ as in equations \eqref{4D7}. As $\theta_{ij}$ and $\theta_{a\,ij}$ are symmetric in $i,j$ one has $\theta_{21}=\theta_{12}$ and $\theta_{a\,21}=\theta_{a\,12}$. Using equations \eqref{4D9} for $\Ns=2$, one easily verifies that all polynomials \eqref{5D8} are cocycles in $\Hg(\fb)$ for $\Ns=2$:
\begin{align}
  \Ns=2:\quad \fb\theta_{a\,ij}=0,\quad \fb\theta_{ij}=0\quad (i,j\in\{1,2\}).
	\label{5D19} 
\end{align}
We are now prepared to prove the following result:

\begin{lemma}[Primitive elements for $\Ns=2$]\label{lem5D1}\quad\\
The general solution of the cocycle condition in $\Hg(\fb)$ for $\Ns=2$ is:
\begin{align}
  \fb\om=0\ \LRA\ \om\sim P(\xi_1,\xi_2)+\theta_{11}P^{11}(\xi_1,\xi_2)+\theta_{12}P^{12}(\xi_1,\xi_2)
  +\theta_{22}P^{22}(\xi_1,\xi_2)\notag\\
                          +\theta_{a\,11}P^{a\,11}(\xi_1,\xi_2)+\theta_{a\,12}P^{a\,12}(\xi_1,\xi_2)
                          +\theta_{a\,22}P^{a\,22}(\xi_1,\xi_2)
	\label{5D20} 
\end{align}
with $\theta_{a\,ij}$ and $\theta_{ij}$ as in equations \eqref{5D8} and arbitrary polynomials $P(\xi_1,\xi_2)$, $P^{a\,ij}(\xi_1,\xi_2)$, $P^{ij}(\xi_1,\xi_2)$ in the components of $\xi_1$ and $\xi_2$.
\end{lemma}

{\bf Proof:} We define the subspace $\hOmg$ of ghost polynomials which do not depend on $c^5$:
\begin{align}
  \hOmg=\Big\{\om\in\Omg\,\Big|\ \frac{\6\om}{\6 c^5}=0\Big\}.
	\label{5D21a} 
\end{align}
Furthermore we define
the subspaces $\Omg^p$ and $\hOmg^p$ of $\Omg$ and $\hOmg$ containing the ghost polynomials with \cdeg\ $p$, respectively:
\begin{align}
  \Omg^p=\{\om\in\Omg\,|\,N_c\, \om=p\,\om\},\quad
  \hOmg^p=\{\om\in\hOmg\,|\ N_c\, \om=p\,\om\},\quad N_c=c^a\frac{\6}{\6 c^a}\,.
	\label{5D21} 
\end{align}
We study the cocycle condition $\fb\om=0$ separately for the various \cdeg s $p$.
As a ghost polynomial is at most linear in $c^5$, each polynomial $\om^p\in\Omg^p$ can be uniquely written as
\begin{align}
  \om^p=c^5\7\om^{p-1}+\7\om^p,\quad \7\om^{p-1}\in\hOmg^{p-1},\ \7\om^p\in\hOmg^p.
	\label{5D22} 
\end{align}
This gives:
\begin{align}
  \fb\om^p=(\fb c^5)\7\om^{p-1}-c^5(\fb\7\om^{p-1})+\fb\7\om^p.
	\label{5D23} 
\end{align}
As $\fb c^5$ is a quadratic polynomial in the supersymmetry ghosts,
only the second term on the right hand side of \eqref{5D23} contains $c^5$. We thus obtain
\begin{align}
  \fb\om^p=0\ \LRA\ (\fb\7\om^{p-1}=0\ \wedge\ (\fb c^5)\7\om^{p-1}+\fb\7\om^p=0).
	\label{5D24} 
\end{align}
Hence, the part $\7\om^{p-1}$ of a cocycle $\om^p$ is a cocycle in $\hOmg$. Furthermore, any contribution $\fb\7\eta^p$ to $\7\om^{p-1}$ with $\7\eta^p\in\hOmg^p$ can be removed from $\om^p$ by adding the coboundary $\fb(c^5\7\eta^p)$ owing to $\om^p+\fb(c^5\7\eta^p)=c^5(\7\om^{p-1}-\fb\7\eta^p)+\7\om^{\prime p}$ with $\7\om^{\prime p}=\7\om^p+(\fb c^5)\7\eta^p\in\hOmg^p$ redefining the part $\7\om^p$ of $\om^p$. Hence, $\7\om^{p-1}$ is actually determined by the cohomology of $\fb$ in $\hOmg$ which we denote by $\hHg(\fb)$. The latter cohomology can be directly obtained from the results in $\Dim=4$ since $\hOmg$ coincides with the $\Dim=4$ space of ghost polynomials and $\fb$ acts identically in both spaces when the identifications \eqref{5D4} are used. In particular, if $\hHg(\fb)$ vanishes at \cdeg\ $p-1$, $\fb\om^p=0$ implies $\om^p\sim\7\om^p$. If additionally $\hHg(\fb)$ also vanishes at \cdeg\ $p$, we obtain $\om^p\sim 0$. Hence, $\Hg(\fb)$ vanishes at \cdeg\ $p$ if $\hHg(\fb)$ vanishes at \cdeg s\ $p-1$ and $p$. Lemma \ref{lem4D8} implies that $\hHg(\fb)$ vanishes for $\Ns=2$ at all \cdeg s $p>1$. We conclude immediately that $\Hg(\fb)$ vanishes at all \cdeg s $p>2$:
\begin{align}
  p>2:\quad \fb\om^p=0\ \LRA\ \om^p\sim 0.
	\label{5D25} 
\end{align}
In the case $p=2$ lemma \ref{lem4D8} implies that the part $\7\om^1$ of $\om^2$ can be taken as
\begin{align}
  p=2:\ \7\om^1=\Theta_{12(\Dim=4)}H(\xi_1,\xi_2)
  +\Theta_{21(\Dim=4)}G(\xi_1,\xi_2)
  +[\Theta_{11}-\Theta_{22}]_{(\Dim=4)}B(\xi_1,\xi_2)
	\label{5D26} 
\end{align}
with polynomials $H$, $G$, $B$ in the components of the supersymmetry ghosts.
Equations \eqref{5D10} to \eqref{5D12} show that $c^5\7\om^1$ can be completed to the cocycle $\theta_{11}P^{11}(\xi_1,\xi_2)+\theta_{12}P^{12}(\xi_1,\xi_2)
  +\theta_{22}P^{22}(\xi_1,\xi_2)$ wherein
\[
P^{11}=-(4k_5)^{-1}G,\ P^{12}=(2k_5)^{-1}B,\
P^{22}(\xi_1,\xi_2)=(4k_5)^{-1}H.
\]
Leaving out the arguments of the $P^{ij}$, this yields for the cocycles $\om^2\in\Omg^2$ the intermediate result
$\om^2=
\theta_{11}P^{11}+\theta_{12}P^{12}
  +\theta_{22}P^{22}+\7\om^{\prime\, 2}$
where $\7\om^{\prime\, 2}\in\hOmg^2$ is the difference of the part $\7\om^2$ in $\om^2=c^5\7\om^1+\7\om^2$ and those terms in $\theta_{11}P^{11}+\theta_{12}P^{12}+\theta_{22}P^{22}$ which do not depend on $c^5$. As $\theta_{11}P^{11}+\theta_{12}P^{12}+\theta_{22}P^{22}$ is a cocycle by itself, the cocycle condition $\fb\om^2=0$ imposes $\fb\7\om^{\prime\, 2}=0$. The latter implies that $\7\om^{\prime\, 2}$ is trivial in $\hHg(\fb)$ since $\hHg(\fb)$ vanishes at \cdeg\ $p=2$ according to lemma \ref{lem4D8}. We conclude in the case $p=2$:
\begin{align}
  \fb\om^2=0\ \LRA\ \om^2\sim \theta_{11}P^{11}(\xi_1,\xi_2)+\theta_{12}P^{12}(\xi_1,\xi_2)
  +\theta_{22}P^{22}(\xi_1,\xi_2).
	\label{5D27} 
\end{align}
The case $p=1$ is somewhat more involved and we shall discuss it without giving all steps in explicit details. In the case $p=1$ the part $\7\om^0$ of a cocycle $\om^1=c^5\7\om^0+\7\om^1$ is purely a polynomial in the components of the supersymmetry ghosts,
\begin{align}
\om^1=c^5\7\om^0(\xi_1,\xi_2)+\7\om^1, \quad \7\om^1\in\hOmg^1.
	\label{5D28} 
\end{align}
Hence, the condition $\fb\7\om^0=0$ in \eqref{5D24} is trivially fulfilled and \eqref{5D24} only imposes $(\fb c^5)\7\om^0+\fb\7\om^1=0$. This yields explicitly in the case $\Ns=2$:
\begin{align}
\Ii(k_5)^{-1}[\xi_1^+\cdot\xi_2^++\xi_1^-\cdot\xi_2^-]_{(\Dim=4)}
\7\om^0(\xi_1,\xi_2)=\fb(-\7\om^1), 
\ \7\om^1\in\hOmg^1.
	\label{5D29} 
\end{align}
Equation \eqref{5D29} imposes that the left hand side of this equation is trivial in $\hHg(\fb)$. Now, any ghost monomial in $\7\om^0$ which depends both on at least one of the components $\xi_i^{+\ua}$ and on at least one of the components $\xi_i^{-\ua}$ yields on the left hand side of equation \eqref{5D29} only terms which are trivial in $\hHg(\fb)$ as in the case $\Ns=2$ one has $2\Ii[(\xi_1^+\cdot\xi_2^+)\xi_2^{-\ua}]_{(\Dim=4)}=\fb\vartheta_{1(\Dim=4)}^{-\ua}$ etc., see equations \eqref{4D9}. In contrast, ghost monomials in $\7\om^0$ which do not depend on any of the components $\xi_i^{+\ua}$ or on any of the components $\xi_i^{-\ua}$ would provide contributions to the left hand side of equation \eqref{5D29} which do not depend on any of the components $\xi_i^{+\ua}$ or on any of the components $\xi_i^{-\ua}$. The latter contributions to the left hand side of equation \eqref{5D29} would not be trivial in $\hHg(\fb)$ because $\fb c^1$, \ldots , $\fb c^4$ only contain monomials which depend both on one of the components $\xi_i^{+\ua}$ and on one of the components $\xi_i^{-\ua}$. This implies that all ghost monomials in $\7\om^0$ depend both on at least one of the components $\xi_i^{+\ua}$ and on at least one of the components $\xi_i^{-\ua}$. Hence, with no loss of generality, $\7\om^0$ can be taken as
\begin{align}
p=1:\ \7\om^0=\sum_{i=1}^2\sum_{j=1}^2\xi_i^{+\ua}\xi_j^{-\ub}\,P_{\ua\ub}^{ij}(\xi_1,\xi_2)
\label{5D30} 
\end{align}
with polynomials $P_{\ua\ub}^{ij}(\xi_1,\xi_2)$ in the components of the supersymmetry ghosts. The sixteen ghost monomials $\xi_i^{+\ua}\xi_j^{-\ub}$ in \eqref{5D30} provide twelve independent cocycles in $\hHg(\fb)$ because $\fb c^1$, \ldots , $\fb c^4$ give four exact linear combinations of these monomials. These twelve cocycles can be taken as the twelve ghost polynomials multiplied by $c^5$ in equations \eqref{5D16} to \eqref{5D18}. This gives, analogously to the case $p=2$, $\om^1\sim \sum_{a=1}^4[\theta_{a\,11}P^{a\,11}(\xi_1,\xi_2)+\theta_{a\,12}P^{a\,12}(\xi_1,\xi_2)
                          +\theta_{a\,22}P^{a\,22}(\xi_1,\xi_2)]
+\7\om^{\prime\, 1}$ with $\7\om^{\prime\, 1}\in\hOmg^1$ and $\fb\7\om^{\prime\, 1}=0$. 
According to lemma \ref{lem4D8}, $\fb\7\om^{\prime\, 1}=0$ implies $\7\om^{\prime\, 1}\sim \Theta_{12(\Dim=4)}H^\prime(\xi_1,\xi_2)
  +\Theta_{21(\Dim=4)}G^\prime(\xi_1,\xi_2)
  +[\Theta_{11}-\Theta_{22}]_{(\Dim=4)}B^\prime(\xi_1,\xi_2)$. By equations \eqref{5D13} to \eqref{5D15} this yields $\7\om^{\prime\, 1}\sim \theta_{5\,11}P^{5\,11}(\xi_1,\xi_2)+\theta_{5\,12}P^{5\,12}(\xi_1,\xi_2)+\theta_{5\,22}P^{5\,22}(\xi_1,\xi_2)$ with $P^{5\,11}=(2k_5)^{-1}G^\prime$, $P^{5\,12}=-(k_5)^{-1}B^\prime$, $P^{5\,22}=-(2k_5)^{-1}H^\prime$. We conclude:
\begin{align}
  \fb\om^1=0\ \LRA\ \om^1\sim \theta_{a\,11}P^{a\,11}(\xi_1,\xi_2)+\theta_{a\,12}P^{a\,12}(\xi_1,\xi_2)
                          +\theta_{a\,22}P^{a\,22}(\xi_1,\xi_2).
	\label{5D31} 
\end{align}
The case $p=0$ is trivial as any element $\om^0$ of $\Omg^0$ is a polynomial $P(\xi_1,\xi_2)$ in the components of the supersymmetry ghosts. Together with \eqref{5D25}, \eqref{5D27} and \eqref{5D31} this proves the lemma.
\QED

{\bf Comment:}
The decomposition \eqref{5D1} of $\Dim=5$ supersymmetry ghosts is not \so{t,5-t}-covariant and, therefore, it was only used in intermediate steps within the derivation of the results in $\Dim=5$ from results in $\Dim=4$. Nevertheless, one may use this decomposition in any particular spinor representation to remove redundant cocycles $P+\theta_{ij}P^{ij}+\theta_{a\, ij}P^{a\, ij}$ in
\eqref{5D20} by restraining the ghost polynomials $P$, $P^{ij}$, $P^{a\,ij}$ analogously to lemma \ref{lem4D8}. For instance, one may always assume that $P^{11}$ and the $P^{a\,11}$ do not depend on the components of $\xi_2^-$, that $P^{22}$ and the $P^{a\,22}$ do not depend on the components of $\xi_1^+$, that $P^{12}$ and the $P^{a\,12}$ do not depend on the components of $\xi_1^+$ or $\xi_2^-$, and that $P$ does not contain terms depending both on components of $\xi_1^+$ and on components of $\xi_2^-$. By refining the proof of lemma \ref{lem5D1} accordingly, this can be deduced directly from lemma \ref{lem4D8} owing to the identifications \eqref{5D4} and analogously for the $P^{a\,ij}$ with $a\neq 5$. Hence, in any particular spinor representation one may specify the result \eqref{5D20} according to:
\begin{align}
  \fb\om=0\ \LRA\ \om\sim \,&P_+(\xi_2^+,\xi_1)+P_-(\xi_1^-,\xi_2)\notag\\
  &+\theta_{11}P^{11}(\xi_2^+,\xi_1)+\theta_{12}P^{12}(\xi_1^-,\xi_2^+)
  +\theta_{22}P^{22}(\xi_1^-,\xi_2)\notag\\
                          &+\theta_{a\,11}P^{a\,11}(\xi_2^+,\xi_1)+\theta_{a\,12}P^{a\,12}(\xi_1^-,\xi_2^+)
                          +\theta_{a\,22}P^{a\,22}(\xi_1^-,\xi_2).
	\label{5D20a} 
\end{align}
We leave it to the interested reader to further specify this result or to characterize the remaining coboundaries along the lines of part (ii) of lemma \ref{lem4D8}.

\subsection{\texorpdfstring{$\Hg(\fb)$ for $\Ns>2$}{H(gh) for N>2}}\label{5D.2}

\begin{lemma}[Absence of primitive elements with \cdeg s $p>1$ for $\Ns>2$]\label{lem5D2}\quad\\
$\Hg(\fb)$ vanishes for $\Ns>2$ at all \cdeg s $p>1$:
\begin{align}
  N_c\,\om^p=p\,\om^p,\ p>1:\quad \fb\om^p=0\ \LRA\ \om^p\sim 0.
	\label{5D32} 
\end{align}
\end{lemma}

{\bf Proof:} As in the proof of lemma \ref{lem5D1} we study the cocycle condition $\fb\om^p=0$ by decomposing it according to equations \eqref{5D24} and by analysing these equations using the results in $\Dim=4$. In the cases $\Ns>2$ lemma \ref{lem4D11} implies that $\hHg(\fb)$ vanishes at all \cdeg s $p>0$. This implies by arguments which led for $\Ns=2$ to the result \eqref{5D25} that $\Hg(\fb)$ vanishes for $\Ns>2$ at all \cdeg s $p>1$.
\QED

{\bf Conjecture:} The author strongly conjectures that $\Hg(\fb)$ vanishes for $\Dim=5$, $\Ns>2$ also at \cdeg\ $p=1$. This would imply that lemma \ref{lem5D2} holds for all \cdeg s $p>0$ in place of $p>1$. 

\section{Conclusion}

We have computed the primitive elements of the supersymmetry algebra cohomology for supersymmetry algebras \eqref{i3-1} in $\Dim=4$ and $\Dim=5$ dimensions, for all signatures $(t,\Dim-t)$, all numbers $\Ns$ of sets of Majorana or symplectic Majorana supersymmetries and all spinor representations equivalent to \eqref{4D1}, except for the particular case of \cdeg\ $p=1$ in $\Dim=5$ for $\Ns>2$ (concerning this case, see the conjecture at the end of section \ref{5D.2}). The results are given in manifestly covariant form in section \ref{4D.2} for $\Dim=4$ (lemmas \ref{lem4D5}, \ref{lem4D8} and \ref{lem4D11}) and in sections \ref{5D.1} and \ref{5D.2} for $\Dim=5$ (lemmas \ref{lem5D1} and \ref{lem5D2}).
We remark that the seemingly preferred role of the supersymmetry ghosts $\xi_1$ in lemmas \ref{lem4D8} and \ref{lem4D11} originates from our method to base the computations in $\Dim=4$ for $\Ns>1$ on the results for $\Ns=1$, and just provides one particular choice of representatives of the cohomology.

As we have explained in some detail in section 7 of \pap,
the results of the present work can be used, inter alia, in the context of algebraic renormalization \cite{Piguet:1995er}, in particular within the classification of counterterms and anomalies, and of consistent deformations \cite{Barnich:1993vg} of supersymmetric (quantum) field theories in four and five dimensions.

\end{document}